\documentclass[final,12pt,twocolumn]{elsarticle}

\usepackage{amssymb}
\usepackage[hyphens]{url}  
\usepackage{amsmath}
\usepackage{xcolor}
\usepackage{multirow}
\usepackage{mathrsfs}
\usepackage{lineno}
\journal{Physica D: Nonlinear Phenomenon}

\newcommand{\avg}[1]{{\left\langle #1 \right\rangle}}

\allowdisplaybreaks

\newif\ifmarkedup
\markedupfalse
\ifmarkedup
\newcommand{\Revision}[2]{\replaced{#2}{#1}}
\else
\newcommand{\Revision}[2]{{\color{black}#2}}
\fi

\newif\ifmarkedupTwo
\markedupTwofalse
\ifmarkedupTwo

\else

\fi
\begin{document}

\begin{frontmatter}



\title{Active Scalar Mixing by Homogeneous Isotropic Turbulence} 


\author{Joaquim P. Jossy}
\author{Pratyush S. Awasthi}
\author{Prateek Gupta\corref{cor}} 
\cortext[cor]{prgupta@iitd.ac.in}

\address{Department of Applied Mechanics, Indian Institute of Technology Delhi, Hauzkhas, New Delhi, 110016, India}

\begin{abstract}
We study the mixing of active scalars by homogeneous isotropic incompressible stochastic velocity fields. We consider both Navier-Stokes generated turbulent fields as well as artificially generated homogeneous isotropic stochastic fields. We use Fourier pseudospectral direct numerical simulations to study the mixing dynamics of two non-reacting species of different density ratios. We use the Atwood number to create a denser mixture and a lighter mixture. We show that in the absence of stirring, a denser mixture homogenizes faster than the lighter mixture. The direction of the density gradient causes the interface across which the molecular diffusion occurs to expand outward for the denser mixture and inward for the lighter mixture. The stirring process, which enhances the diffusion process, increases the rate of homogenization in both mixing methods under study. We define a new mixing metric for studying the mixing evolution of active scalars, which indicates that a denser inhomogeneity in a lighter mixture spreads faster but homogenizes slower. For low Mach number turbulence,  there is a negligible coupling between the density gradients and the velocity field responsible for stirring. The post-stirring behavior of active scalars is found to be similar to passive scalars, where the scalar energy spectra decay exponentially and exhibit self-similarity. The turbulence fields generated by solving the Navier-Stokes equation homogenize both the mixtures faster than the synthetic cases. We show that matching the kinetic energy spectra and inertial subrange scaling of a synthetically generated stochastic field with that of a Navier-Stokes generated field is not enough to study mixing dynamics.
\end{abstract}




\end{frontmatter}



\section{Introduction}
\Revision{}{The evolution of an initially segregated mixture into a homogenized state is called mixing. Mixing can be classified into three types, depending on the effect of the constituents of the mixture on the kinematics~\cite{dimotakis2005turbulent}}. The mixing process in which the constituents do not affect the flow-kinematics is called the level-1 mixing process. Mixing of passive scalars such as smoke, dyes, pollutants, etc., is an example of level-1 mixing. In level-2 mixing, the constituent density fields and the flow velocity are coupled to each other. Some examples are Ritchmyer-Meshkov driven mixing, Rayleigh-Taylor instability mixing, compressible turbulence, and mixing driven by random shocks \cite{sandoval1995dynamics, livescu2010new, thornber2012energy, ni2016compressible, zhou2017rayleigh, livescu2020turbulence, bender2021simulation, jossy2023baroclinic, jossy2023spectral, zhou2024hydrodynamic, zhou2025instabilities, jossy2025mixing}. Level-3 mixing differs from level-2 mixing in that the composition of the constituents changes in level-3 mixing while it remains constant in level-2 mixing. Combustion is one of the examples of level-3 mixing. In this work, we focus on level-2 mixing driven by a \Revision{random}{stochastic} velocity field, where two non-reacting species of different densities are mixed together. Understanding active scalar mixing has a wide range of engineering applications -  from the development of inertial confinement fusion reactors \cite{zhou2025instabilities}, combustion chambers design \cite{yang1993applications}, building stellar evolution models \cite{arnett2000role}, and predicting atmospheric flows.

The mixing process is governed by two actions, molecular diffusion and stirring. The molecular diffusion process smoothens out concentration gradients, and its action is reflected through the Laplacian term in the concentration evolution equation. \Revision{The advection term in the evolution equation denotes the stirring action}{The action of the advection term was referred as ``stirring" by~\citeauthor{eckart1948analysis}~\cite{eckart1948analysis}}. While complete homogenization can be achieved by molecular diffusion alone, stirring increases the rate of homogenization. Stirring enhances the diffusion process by increasing the surface area across which the molecular diffusion occurs. This is done by stretching and folding of the interface. In active scalar mixing, the presence of density gradients alters this stirring process. \citeauthor{jossy2025mixing}~\cite{jossy2025mixing} show through two-dimensional direct numerical simulations (DNS) that the nonlinear dissipation terms consisting of the density gradient alter the stirring process. A similar breakdown of the concentration diffusion term was also shown by Yu et al.~\cite{yu2021scaling} in shock bubble interaction studies. Such nonlinear dissipation causes an apparent movement of the inhomogeneity interface. The interaction of stirring dynamics with Laplacian diffusion and this interface movement results in different regimes throughout the mixing process. Gonzalez and  Parantho\"en
\cite{gonzalez2011effect} analytically study the effect of density gradients on the orientation and on the norm of the concentration gradient. They show that the density gradients affect the kinematics of the flow field through the velocity divergence. In this work, we investigate the additional effects needed to model the \Revision{}{effect of mixing on kinematics} using direct numerical simulations of \Revision{}{homogeneous isotropic turbulent fields and stochastically generated velocity fields, also called as ``synthetic turbulent fields"~\cite{careta1993stochastic}}. 

The mixing enhancement \Revision{of}{induced by} stirring is primarily attributed to \Revision{its capability of generating and maintaining}{the continuous generation of} vortical structures. Previous studies have shown that compressibility of the flow has a minimal effect on enhancing the mixing process \cite{jossy2025mixing,ni2016compressible,panickacheril2019solenoidal}. Meunier and Villermaux \cite{meunier2003vortices} study the mixing of a passive scalar blob using a Lamb-Oseen vortex in two dimensions. They show the action of stirring exists till points of the blob are exposed to the action of molecular diffusion.
\citeauthor{careta1994diffusion} \cite{careta1994diffusion} study the effective diffusion of passive scalars using synthetically generated homogeneous isotropic stochastic velocity fields. \citeauthor{toussaint2000spectral}~\cite{toussaint2000spectral} study the mixing of passive scalars using two-dimensional recirculating flows. They report that the decay phase of the scalar spectrum (rate of homogenization) is different for different kinds of stirring action. \citeauthor{juneja1996dns}~\cite{juneja1996dns} study the mixing of two passive scalars in stationary, homogeneous, isotropic turbulence using direct
numerical simulations. They show that in the decay phase, the scalar fields exhibit self-similar behavior. \citeauthor{orsi2021scalar}~\cite{orsi2021scalar} study the mixing of passive scalars using a statistically stationary field of homogeneous, isotropic turbulence. They show that the gamma distribution is a suitable model for the distribution function of the passive scalar concentration in space. \citeauthor{schumacher2005statistics} \cite{schumacher2005statistics} derive a relation between the geometrical properties
and the mixing statistics of passive scalars when stirring is done by a stochastically forced isotropic turbulent field. They show that passive scalar concentration fluctuations follow a Gaussian distribution by relating scalar concentration to the changes in area and volume. \citeauthor{livescu2008variable}~\cite{livescu2008variable} study the mixing of binary mixture where the stirring is induced by buoyancy-generated motion.  They conclude that heavier fluids undergo less stirring than lighter fluids when buoyancy is used as the stirring action. In our previous work \cite{jossy2025mixing}, we have reported similar results when a binary mixture is stirred using a field of random shock waves in two dimensions, with baroclinicity being the cause of vorticity. The field of random shock waves has a broadband spectrum with -2 slope~\cite{gupta2018spectral, jossy2023baroclinic, thomas2024turbulent, jossy2025mixing}. Hence, the vortical field is generated at a multitude of length scales. We have shown that the denser mixtures undergo more nonlinear dissipation than stirring, resulting in a delayed effective action of molecular diffusion. In this work, we use direct numerical simulations to study the mixing of binary miscible mixtures of varying densities when they are stirred using a homogeneous isotropic field. Though a lot of studies exist on the mixing of passive scalars, very little literature exists on comparing the behavior of passive scalars with active scalars. In this work, we compare the mixing characteristics of active and passive scalars  by stochastic velocity fields generated by solving the Navier-Stokes equations and by those generated synthetically~\cite{careta1993stochastic,holzer1994turbulent,rosales2011synthetic}. We also compare the differences in stirring by the two different stochastic fields.

\Revision{}{We consider binary mixtures of two non-reacting species with different density ratios. Throughout, we consider a spherical inhomogeneity (blob) and vary the densities to obtain lighter and denser mixtures. In the next section, section \ref{sec: Governing equations}, we outline the dimensionless governing equations used for simulations and analysis in this work. In the next section, section~\ref{sec: 3D-DNS} we discuss the numerical methodology used for 3D DNS of mixing of active scalars due to a very low Mach number (incompressible) homogeneous isotropic turbulent (HIT) flow field. Further in section~\ref{sec:3D Synthetic Velocity Fields}, we discuss the generation of synthetic stochastic velocity fields to understand the mixing dynamics of active scalars due to a synthetically generated field. In section \ref{sec: Results}, we compare the mixing difference between active and passive scalars. We also highlight the differences in the mixing process for the two different mixing methods (3D DNS and synthetic fields), before concluding in the last section.}

\Revision{}{\section{Governing equations for a gaseous mixture}
\label{sec: Governing equations}
Equations governing the mixing of two gaseous species are the compressible Navier-Stokes equations combined with species equation and the equation of state (assumed to ideal gas in this work). In the dimensionless form, these equations for density $\rho$, velocity $\boldsymbol{u}$, pressure $p$, and mass-fraction of the $i^{\mathrm{th}}$ species $Y_i$ are given by, 
\begin{subequations}
    \begin{align}
    &\frac{\partial \rho}{ \partial t} +  \nabla\cdot(\rho \boldsymbol{u}) =0 ,
    \label{eq: continuity}\\
    &\frac{\partial \boldsymbol{u} }{\partial t} + \boldsymbol{u} \cdot (\nabla \boldsymbol{u}) + \frac{\nabla p}{\rho} = \frac{1}{ \rho \mathrm{Re} }  \left( \nabla \cdot ( 2 \mu \boldsymbol{S} )\right) \nonumber \\
    & + \frac{1}{ \rho \mathrm{Re} }  \left( \nabla  \left( \kappa - \frac{2 \mu }{3} \right) \nabla\cdot\boldsymbol{u} \right) + \boldsymbol{F} ,
    \label{eq:momentum_equation}\\
    &\frac{\partial p}{\partial t} + \boldsymbol{u} \cdot \nabla p + \gamma p \nabla \cdot \boldsymbol{u} =  
    \frac{1}{M^2 \mathrm{Re}  \mathrm{Pr}} \nabla \cdot (\alpha \nabla T)  \nonumber \\
    & +\frac{\gamma -1 }{\mathrm{Re}} \left( 2 \mu \boldsymbol{S}:\boldsymbol{S}  
    + \left( \kappa - \frac{2 \mu}{3} \right)\left( \nabla\cdot\boldsymbol{u}\right)^2 \right) ,
    \label{eq:pressure_equation}\\
    &\frac{\partial Y_{i}}{\partial t} + \boldsymbol{u} \cdot \nabla Y_{i} = \frac{\nabla \cdot \left( \rho  \nabla Y_{i}\right)}{\rho  \mathrm{Re}  \mathrm{ Sc}},
     \label{eq: species_gov_equation}
    \end{align}
    \label{eq: gov_eqs}
\end{subequations}
where $\boldsymbol{F}$ is an external force, $M$ is a reference Mach number, and $\gamma=1.4$ is the adiabatic index. In the species evolution equation, we assume a constant effective diffusion coefficient, resulting in the Fick's law for species density~\cite{poinsot2005theoretical}. We close the system of equations~\eqref{eq: gov_eqs} using the dimensionless equation of state for an ideal gas mixture, 
\begin{equation}
\gamma M^2 p = \frac{\rho}{W} T,\label{eq: ideal_gas_nd}        
\end{equation}where $W$ is the mean molecular mass of the mixture and is given by 
\begin{equation}
    \frac{1}{W} = \frac{Y_{1}}{W_{1}} + \frac{Y_{2}}{W_{2}}.
        \label{eq: mean molecular weight equation}
\end{equation}
In \eqref{eq:momentum_equation} and \eqref{eq:pressure_equation}, $\boldsymbol{S}$ is the strain rate tensor given by, 
\begin{equation}
\boldsymbol{S} = \frac{1}{2}\left(\nabla\boldsymbol{u} + \nabla\boldsymbol{u}^T\right). \label{eq:strain_tensor}
\end{equation}
 The fluid properties such as dynamic viscosity ($\mu$), bulk viscosity ($\kappa$), and thermal conductivity ($\alpha$) are all dimensionless. The flow properties are governed by the dimensionless numbers, the Reynolds number ($\mathrm{Re}$), the Mach number ($M$), the Prandtl number ($\mathrm{Pr}$), and the Schmidt number ($\mathrm{Sc}$) defined as,
\begin{subequations} \label{eq:non_dimensional_parameters}
\begin{align}
    \mathrm{Re} = \frac{\rho_{m} u_{m} L_{m}}{\mu_{m}}, &~ M =\frac{u_m}{c_m}\\[5pt]
    \mathrm{Pr} = \frac{\gamma \mu_{m} \tilde{R}}{(\gamma - 1) W_{m} \alpha_{m}}, &~
    \mathrm{Sc} = \frac{\mu_{m}}{\rho_{m} \lambda_{m}},
\end{align}
\end{subequations}
where $()_m$ denotes reference dimensional quantities used for non-dimensionalization. The Reynolds number in \eqref{eq:momentum_equation} and \eqref{eq:pressure_equation} is calculated based on characteristic dimensional quantities of velocity, length, viscosity, and density~\cite{zang1992direct} (hereafter referred as the computational Reynolds number). $\tilde{R}$, $\mu_{m}, \alpha_{m} $, and $\lambda_{m}$ denote the universal gas constant, dimensional viscosity, thermal conductivity, and diffusion coefficient, respectively. Since we use dimensionless equations in our numerical simulations, the specific values of the characteristic scales hold no significance. We also assume that the heat capacities of both the species are same and constant in temperature.  Finally, $c_m = \sqrt{\gamma p_m/\rho_m}$ is the speed of sound at reference state. 

\section{3D DNS of mixing due to incompressible HIT}
\label{sec: 3D-DNS}
To study mixing due to incompressible HIT, we set the Mach number to a low value ($M=0.1$) in all our simulations, thus ensuring that the flow field is predominantly incompressible. 
We use an MPI parallelized Fourier pseudo-spectral which uses the P3DFFT library~\cite{pekurovsky2012p3dfft} to study the mixing of active scalars by both Navier-Stokes generated turbulent fields as well as artificially generated homogeneous isotropic stochastic fields. We use a fourth-order Runge-Kutta explicit time-stepping scheme for time integration. For all the simulations discussed in this work, we consider a triply periodic box of side-length $L=2\pi$, and consider 192 Fourier modes in each direction. We use the standard $2/3$ de-aliasing method~\cite{jossy2023baroclinic, thomas2024turbulent, jossy2025mixing} such that the nonlinear terms are resolved till $k =64$. We choose Reynolds number of $\mathrm{Re} = 125$ to achieve a fully resolved simulation within available resources, and $\mathrm{Pr} = \mathrm{Sc} = 0.7$.

\subsection{Stochastic Forcing}
To generate the turbulent flow field, we use a stochastic forcing following \citeauthor{eswaran1988examination}~\cite{eswaran1988examination}. We generate homogeneous isotropic turbulence in a three-dimensional box by a stochastic forcing using the Uhlenbeck-Ornstein (UO) process \cite{eswaran1988examination}. The forcing $\boldsymbol{F}$ in \eqref{eq:momentum_equation} is defined in the Fourier space using six independent solutions of a UO process $a_{i,\boldsymbol{k}}(t)$ for $i=1, 2, 3, 4, 5$ and $6$ (one complex number in each direction) whose evolution is obtained by,
\begin{equation}
    a_{i,\boldsymbol{k}}(t+dt) = a_{i,\boldsymbol{k}}(t) (1 - dt/T_L) + \boldsymbol{\chi} (2 \sigma^2 dt/T_L)^{1/2}
    \label{eq: Solving UO}
\end{equation}
where  $\sigma^2$ denotes the variance, $T_L$ is the forcing time-scale, and  $\boldsymbol{\chi}$ is a delta correlated Gaussian vector of zero mean. Combining independent processes $a_1, a_2, a_3$, $a_4$, $a_5$ and $a_6$, the forcing vector in the Fourier space can be obtained as, 
\begin{equation}
    \widehat{\boldsymbol{a}}_{\boldsymbol{k}} = \left(a_{1,\boldsymbol{k}} + i a_{2,\boldsymbol{k}}, a_{3,\boldsymbol{k}} + ia_{4,\boldsymbol{k}} , a_{5,\boldsymbol{k}} + ia_{6,\boldsymbol{k}} \right)^\mathscr{T}.
    \label{eq: randomForcingVector}
\end{equation}
where $\mathscr{T}$ denotes transpose. The evolution of each UO process defined by equation \ref{eq: UO process} has zero mean and has time correlation which exponentially decay. Thus following conditions are imposed on each of the $a_{i,\boldsymbol{k}}$,
\begin{align}
 &\left\langle a_{i, \boldsymbol{k}}(t)\right\rangle = 0 ,\\
  &\left\langle {a}_{i, \boldsymbol{k}}(t) {a}_{j, \boldsymbol{k}}^{*}(t+s)\right\rangle = 2 \sigma^{2} \delta_{ij}\exp(-s/T_L) ,
\end{align}
where $\left\langle\cdot\right\rangle$ denotes the ensemble average, $()^*$ denotes the complex conjugate. We restrict $\boldsymbol{F}$ to a band of wavenumbers $0<|\boldsymbol{k}|<k_F$ to ensure only large-scale forcing. The total rate of energy addition $\epsilon$ in the system depends on the variance $\sigma^2$, time-scale $T_L$, and number of wavenumber vectors within the forcing band $N_F$ as, 
\begin{align}
  \epsilon = 4N_F T_L \sigma^{2}.
  \label{eq: energyAddition}
\end{align}
In all our 3D DNS presented in this work, we use the following values of forcing parameters: $T_L = 0.01$, $\sigma^2  = 0.5 $, and $k_F =2\sqrt{2}$.
\subsection{Initial condition of an inhomogeneous mixture in a turbulent field}
We specify the species having a spherical distribution using a $\tanh$ function as follows
\begin{equation}
    Y_{c} = \frac{1}{2} \Bigg[ 1 - \mathrm{tanh} \Bigg( \frac{1}{\delta} 
    \Bigg( S  - \frac{\pi}{2} \Bigg) \Bigg) \Bigg] .
\label{eq: circular species distribution}
\end{equation}
\begin{equation}
   S = \sqrt{(x - \pi)^2 + (y - \pi)^2 + (z - \pi)^2}
\end{equation}
The species constituting the blob has density $\rho_{c}$, and the surrounding species has density $\rho_{s}$. We use the subscript $()_c$ to indicate the species inside the spherical blob and the subscript $()_s$ to denote the surrounding species. We use the parameter value $\delta = 1/5$ in \eqref{eq: circular species distribution} (approximately more than 10 times the maximum scalar gradient possible) to obtain a steep but smooth concentration distribution. To characterize different density ratios, we use the Atwood number ($At$) defined as 
\begin{equation}
   At = \frac{\rho_{s} - \rho_{c}}{\rho_{s} + \rho_{c}} \hspace{1 mm} = \frac{\chi - 1}{\chi + 1} .
\end{equation}
We study the effect of stirring on both a denser mixture ($\chi > 1$) and a lighter mixture ($\chi < 1$), where $\chi$ is the density ratio of the surrounding species to the blob species.

To prevent transient effects of developing homogeneous isotropic turbulence on the inhomogeneous mixture, we initialize the velocity and pressure fields in the low Mach number compressible simulations with a statistically stationary velocity and pressure field obtained from an incompressible DNS to avoid any transient effects. This ensures that the mixing dynamics reported are purely due to an established turbulent velocity field. The incompressible DNS is run with the identical forcing discussed in the previous section with the same forcing parameters ($T_L = 0.01$, $\sigma^2  = 0.5 $, and $k_F =2\sqrt{2}$)till the statistically stationary state is achieved. To initialize the low Mach number compressible simulations using the incompressible HIT, we follow the asymptotic expansion given by~\citeauthor{ristorcelli1997consistent}~\cite{ristorcelli1997consistent}. We initialize the pressure field $p$ as, 
\begin{equation}
    p = \frac{1}{\gamma M^2} + p_*,
    \label{eq: comp_pres_from_incomp_pres}
\end{equation}
where $p_*$ is the pressure field of the incompressible simulation at statistically stationary state. The velocity field is initialized as $\boldsymbol{u} = \boldsymbol{u}_*$, where $\boldsymbol{u}_*$ denotes the velocity field of the incompressible simulation at statistically stationary state. Using the initial pressure from \eqref{eq: comp_pres_from_incomp_pres} and molecular weight from \eqref{eq: circular species distribution} and \eqref{eq: mean molecular weight equation} in the equation of state \eqref{eq: ideal_gas_nd}, along with isothermal conditions ($T=1$), we initialize the non-uniform density. Figure~\ref{fig: Kinetic Energy} shows the evolution of the mean kinetic energy per unit mass as the low Mach number compressible DNS are initialized from an incompressible DNS for all the parameters discussed below.

\subsection{Simulation parameters}
We run five cases of 3D DNS using the compressible solver. Three of these simulations are simulated with different spatially averaged density values denoted by $\overline{\rho}$ (labeled using ``b'' in table~\ref{tab: DNS-cases}), and the remaining two cases are initialized by keeping equal spatially averaged density (denoted by $\overline{\rho}$) but with different density ratios (labeled using $d$ in table~\ref{tab: DNS-cases}). 
\begin{table}[!t]
\centering
\begin{tabular}{|l|c|c|}
\hline
 \multicolumn{3}{|c|}{\textbf{3D DNS}} \\ \hline
                 &{Different $\overline{\rho}$}&{Same $\overline{\rho}$}\\ \hline
${At = 0.75}$       & 1b ($\overline{\rho}$ = 6.43) & 1d ($\overline{\rho}$ = 1.00) \\ \hline
${At = -0.75}$      & 2b ($\overline{\rho}$ = 0.18) & 2d ($\overline{\rho}$ = 1.00) \\ \hline
{Passive}           & \multicolumn{2}{c|}{3b ($\overline{\rho}$ = 1.00)  }  \\ \hline
\end{tabular}
\caption{Parameter space for 3D DNS and their indicators.}
\label{tab: DNS-cases}
\end{table}
\begin{table}[!b]
\centering
\begin{tabular}{|c|c|c|c|c|}
\hline
\textbf{\textbf{Cases}}
&\textbf{\textbf{$\eta$}}
&\textbf{$k_{max}\eta$} &\textbf{$\eta_B/\Delta$}
&\textbf{\textbf{$Re_{\lambda}$}}\\ \hline
$*$  & 0.02  & 1.7  & - & 63 \\ \hline
${1b}$  & 0.0068  & 0.44 & 0.174 & 138 \\ \hline
${2b}$  & 0.0975  & 6.24  & 2.504 & 15  \\ \hline
${3b}$  & 0.0269 & 1.73 & 0.693 & 48 \\ \hline
${1d}$  & 0.0272  & 1.74 & 0.698 & 49 \\ \hline
${2d}$  & 0.0293 & 1.88 & 0.755 & 46  \\ \hline
\end{tabular}
\caption{Resolution quantities of the DNS cases, where $\eta$ is the time averaged Kolmogrov scale, $\eta_B$ is the time averaged Batchelor scale and $Re_\lambda$ is the time averaged Taylor Reynolds number.}
\label{tab: resolved quantities}
\end{table}
To ensure that the DNS are resolved, we calculate the Kolmogorov scale $(\eta)$, following  \citet{wang2017shocklet}. The time-averaged $\eta$  and  $k_{max} \eta$ values for each of the DNS cases are shown in Table \ref{tab: resolved quantities}.  For the chosen energy injection rate, the time-averaged $k_{max} \eta$ is greater than 1.5  for all cases except for the denser mixture, with different mean density (case 1b). The comparison between denser and lighter mixtures for the rest of the study will be done using the cases where $\overline{\rho} = 1$ (cases 1d and 2d). We use the different mean density cases to show that the results hold irrespective of the mean density value. The corresponding time-averaged Taylor microscale Reynolds number $Re_{\lambda}$ is also shown in Table \ref{tab: resolved quantities}.  The reason for the generation of more scales in positive Atwood numbers and the drop in scales in negative Atwood numbers with different mean densities will be discussed in Section \ref{sec: Effect of density gradients on velocity}. To ensure the resolution of the active scale, we calculate the smallest scale for concentration gradients, the Batchelor scale $(\eta_B = \eta / \sqrt{\mathrm{Sc}})$. We follow an even more strict resolution scale for mixing, $   \eta_{B}/\Delta > 0.5 $~\cite{schumacher2005statistics}
where $\Delta$ is given by $ 2\sqrt{2} \pi / 3k_{\mathrm{max}}$. The corresponding values for each of the mixing DNS cases are also shown in Table \ref{tab: resolved quantities}.
\begin{figure}[!t]
    \centering
    \includegraphics[width=1.0\linewidth]{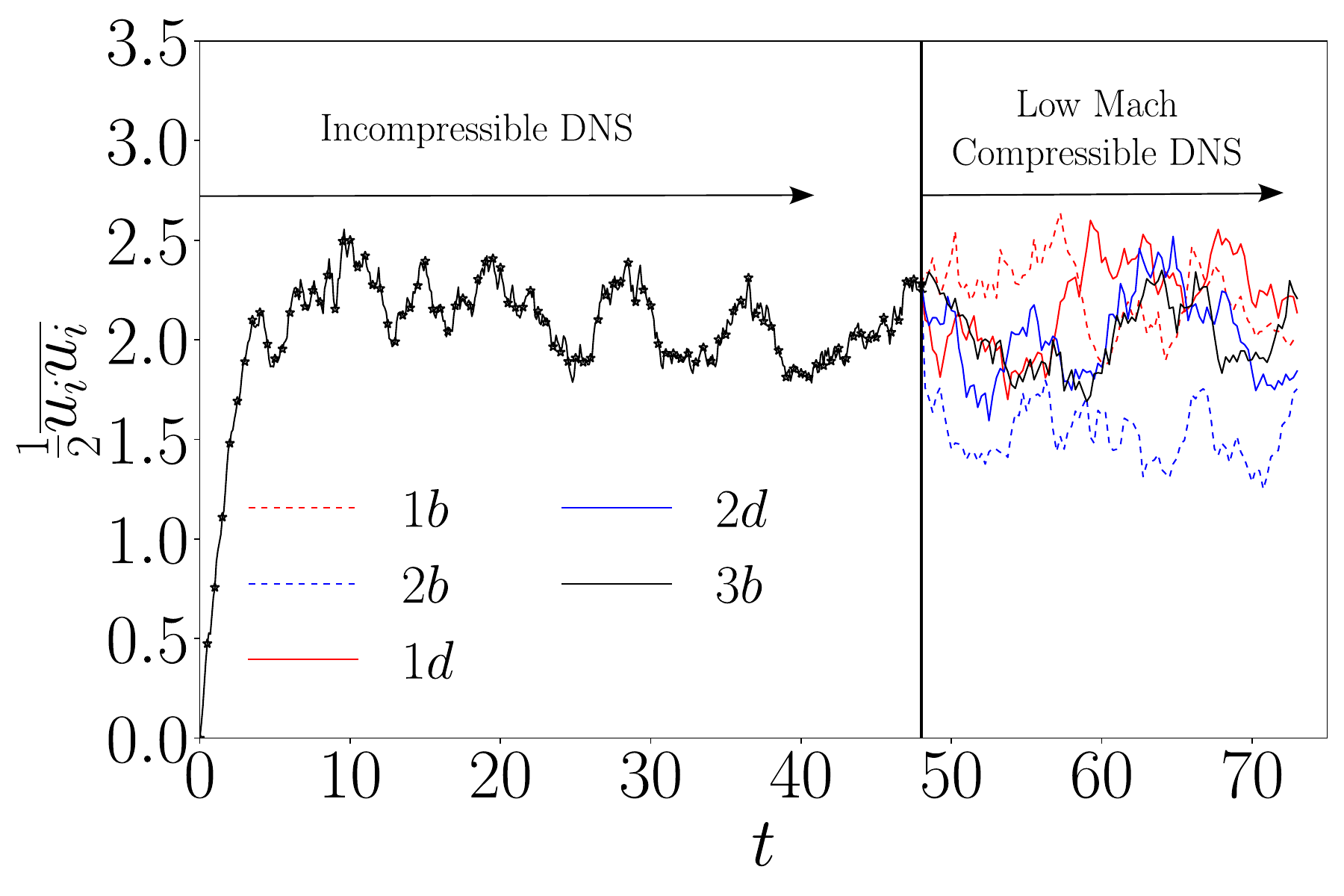}
    \caption{Evolution of the spatially averaged kinetic energy for incompressible DNS and $At=0.75$, $At=-0.75$, and passive scalar for low Mach compressible 3D DNS.}
    \label{fig: Kinetic Energy}
\end{figure}
}

\section{ 3D Stochastic Velocity Fields}
\label{sec:3D Synthetic Velocity Fields}
Turbulence induces motion of the inhomogeneities in the mixture at various length scales. To model the mixing of an inhomogeneous mixture by turbulence, we also study the mixing of active scalars by synthetically generated stochastic fields. We use the governing equation for density similar to those used in earlier variable density studies \cite{livescu2008variable,sandoval1995dynamics,livescu2010new,baltzer2020variable,livescu2020turbulence}. For the incompressible flow of a binary mixture of two non-reacting species,  the dimensionless density of the mixture $\rho$ becomes a function of only the mass fraction of the species ($Y_i$) and is given by 
\begin{equation}
    \frac{1}{\rho} = \frac{Y_{1}}{\rho_{1}} + \frac{Y_{2}}{\rho_{2}}.
        \label{eq: mean molecular weight equation incompressible}
\end{equation}
The above equation is valid only for isobaric and isothermal conditions as long as the microscopic densities of the two species remain constant. Additionally, in a variable density incompressible flow, the velocity field $\boldsymbol{u}$ is constrained by~\cite{ramshaw2002fluid}
\begin{equation}
    \nabla \cdot \boldsymbol{u} = - \frac{1}{\mathrm{Re} \mathrm{Sc}} \nabla \cdot \left(\frac{\nabla \rho}{\rho}\right),
    \label{eq: velocity constraint}
\end{equation}
yielding the following dimesionless density equation
\begin{equation}
    \frac{\partial \rho}{ \partial t} +    \boldsymbol{u}_c \cdot \nabla \rho = \frac{\rho}{\mathrm{Re}  \mathrm{Sc}} \nabla^2\rho ,
    \label{eq: density}
\end{equation}
with the species conservation equation remaining the same as \eqref{eq: species_gov_equation}. Note that in \eqref{eq: density}, the velocity $\boldsymbol{u}_c$ accounts for the additional velocity field (purely vortical in nature) which may be synthetically generated or obtained from a Navier-Stokes simulation. The action of stirring is accounted for by the term $\boldsymbol{u}_c\cdot\nabla \rho$. Previous studies show that the vortical velocity component is responsible for the stretching and the folding of the concentration interface. 
We follow \citeauthor{careta1993stochastic}~\cite{careta1993stochastic} and use stochastic differential equations to generate divergence free velocity fields $\boldsymbol{u}(\boldsymbol{r},t)$, which have zero mean, are homogeneous and isotropic, and exhibit a pre-determined spectra. To impose the divergence-free condition naturally, we use a vector potential $\boldsymbol{\eta}$ which is generated using the Uhlenbeck-Ornstein process give by,
\begin{equation}
    \tau \frac{\partial \boldsymbol{\eta} (\boldsymbol{r},t)}{\partial t} = -\boldsymbol{\eta}(\boldsymbol{r},t) + \mathcal{Q}\boldsymbol{\chi}(\boldsymbol{r},t).
    \label{eq: UO process}
\end{equation}
where $\boldsymbol{\chi}$ is a delta correlated Gaussian vector of zero mean with its correlation tensor defined as
\begin{equation}
    \left\langle \chi_i\left(\boldsymbol{r}_{1},t_{1} \right)\chi_j\left(\boldsymbol{r}_{1},t_{2}\right)\right\rangle = 2 \varepsilon \delta_{ij}\delta(\boldsymbol{r}_1 -\boldsymbol{r}_2) \delta(t_1 -t_2),
\end{equation}
where $\varepsilon$ represents the intensity of the generated white noise. The operation $Q\boldsymbol{\chi}$ is defined such that the Fourier transform of $Q\boldsymbol{\chi}$ is the product $\hat{Q}\hat{\boldsymbol{\chi}}$, where $\hat{Q}$ is the filtering kernel. Throughout, we use $\hat{()}$ to denote the Fourier transform. We assume that all the fields exist in a box $B$ of side-length $2\pi$ and vanish outside the box, and hence define the Fourier transform for a field $\phi(\boldsymbol{x})$ as,
\begin{equation}
\hat{\phi}(\boldsymbol{k}) = \frac{1}{(2\pi)^3}\int\displaylimits_B \phi(\boldsymbol{x})e^{-i\boldsymbol{k}\cdot\boldsymbol{x}}d\boldsymbol{x} = \overline{\phi(\boldsymbol{x})e^{-i\boldsymbol{k}\cdot\boldsymbol{x}}},
\end{equation}
where $\overline{()}$ denotes spatial average in the box. The reader is referred to~\citeauthor{batchelor1953theory}~\cite{batchelor1953theory} for a discussion on the existence and convergence of the inverse relation integral. The filtering kernel $\hat{Q}$ is used to achieve the desired variation of the energy spectra. The solution of the vector stream function in Fourier space is given by 
\begin{equation}
    \hat{\boldsymbol{\eta}} (\boldsymbol{k},t) = \hat{\boldsymbol{\eta}} (\boldsymbol{k},0) e^{-t/\tau} + \frac{1}{\tau} \int\displaylimits_{0}^{t} e^{\frac{t'-t}{\tau} } \hat{\mathcal{Q}}\hat{\boldsymbol{\chi}} dt'
    \label{eq: Solution of Stochastic eqaution}
\end{equation}
The velocity field is computed using $\boldsymbol{\eta}$ as $\boldsymbol{u} = \nabla\times \boldsymbol{\eta}$. The relation between the auto-correlation of $\hat{\boldsymbol{\eta}}(\boldsymbol{k},t)$ and that of $\hat{\boldsymbol{u}}(\boldsymbol{k},t)$ is given by, 
\begin{equation}
    \avg{\hat{\boldsymbol{u}}^*(\boldsymbol{k})\cdot\hat{\boldsymbol{u}}(\boldsymbol{k})} = k^2\avg{\hat{\eta}^*_q\hat{\eta}_q} - k_l k_p\avg{\hat{\eta}^*_l\hat{\eta}_p},
    \label{eq: uhat_etahat_correlation}
\end{equation}
for all time pairs, where $k = |\boldsymbol{k}|$. From \eqref{eq: Solution of Stochastic eqaution}, we obtain,
\begin{align}
&\avg{\hat{\eta}^*_p(\boldsymbol{k},t)\hat{\eta}_q(\boldsymbol{k},t+s)} \nonumber \\ &= \hat{Q}^*\hat{Q}\frac{e^{-s/\tau}}{\tau^2}\int\displaylimits^t_0\int\displaylimits^{t+s}_0\avg{\hat{\chi}^*_p(t')\hat{\chi}_q(t'')}e^{\frac{t'+t'' - 2t}{\tau}}dt'dt''\nonumber \\ 
& = \hat{Q}^*\hat{Q}\frac{2\varepsilon \delta_{pq}e^{-s/\tau}}{(2\pi)^3\tau^2}\int\displaylimits^t_0\int\displaylimits^{t+s}_0\delta(t'-t'')e^{\frac{t'+t'' - 2t}{\tau}}dt'dt''.
\end{align}
In the limit $t\to\infty$, the above expression yields, 
\begin{equation}
\lim_{t\to\infty}\avg{\hat{\eta}^*_p(\boldsymbol{k},t)\hat{\eta}_q(\boldsymbol{k},t+s)} = 2\varepsilon\delta_{pq}\frac{\hat{Q}^*\hat{Q}}{(2\pi)^3\tau}e^{-s/\tau}.
\label{eq: eta_correlation_result}
\end{equation}
Substituting \eqref{eq: eta_correlation_result} in \eqref{eq: uhat_etahat_correlation}, we obtain, 
\begin{align}
\lim_{t\to\infty}\avg{\hat{\boldsymbol{u}}^*(\boldsymbol{k}, t)\cdot\hat{\boldsymbol{u}}(\boldsymbol{k}, t+s)}=
k^2\varepsilon\frac{\hat{Q}^*\hat{Q}}{2\pi^3\tau}e^{-s/\tau}.
\label{eq: uhat_Qhat}
\end{align}
Additionally, using the definition of $\hat{\boldsymbol{u}}$ we can show, 
\begin{align}
&\avg{\hat{\boldsymbol{u}}^*(\boldsymbol{k}, t)\cdot\hat{\boldsymbol{u}}(\boldsymbol{k}, t+s)} =\nonumber \\
&\frac{1}{(2\pi)^3}\int\displaylimits_B\frac{d\boldsymbol{x}}{(2\pi)^3}\int\displaylimits_B\avg{\boldsymbol{u}(\boldsymbol{x},t)\cdot\boldsymbol{u}(\boldsymbol{x+r,t+s)}}\nonumber\\
&e^{-i\boldsymbol{k}\cdot\boldsymbol{r}}d\boldsymbol{r}.
\end{align}
Since we are generating a homogeneous field, the velocity correlation inside the integrand is independent of $\boldsymbol{x}$. The remaining expression shows that $\avg{\hat{\boldsymbol{u}}^*(\boldsymbol{k}, t)\cdot\hat{\boldsymbol{u}}(\boldsymbol{k}, t+s)}$ is the Fourier transform of the velocity auto-correlation function $R_{ii}(\boldsymbol{r},s)$ defined using,
\begin{equation}
R_{ij}(\boldsymbol{r},s) = \avg{u_i(0, t)u_j(\boldsymbol{r},t+s)}.
\label{eq: Radial correlation define}
\end{equation}
Additionally, for homogeneous isotropic turbulence, the Fourier transform of the velocity auto-correlation $\phi_{ii}(\boldsymbol{k},0)$ is related to the energy spectra $E(k)$ as~\cite{batchelor1953theory},
\begin{equation}
    \phi_{ii} = \frac{E(k)}{2\pi k^2}.
\end{equation}
Using \eqref{eq: uhat_Qhat}, we obtain, 
\begin{equation}
E(k) = k^4 \varepsilon\frac{\hat{Q}^*\hat{Q}}{\pi^2\tau}.
\label{eq: Ek_Qhat_relation}
\end{equation}
Using the radial auto-correlation function of velocity $R(r,s) = \frac{1}{3}R_{ii}(\boldsymbol{r},s)$ for isotropic turbulence and its relation with $E(k)$, the correlation for the synthetically generated field can be evaluated as, 
\begin{equation}
R(r,0) = \frac{2}{3}\int\displaylimits^\infty_0E(k)\frac{\sin(kr)}{kr}dk.
\label{eq: R_Ek_relation}
\end{equation}
The choice of the spatial filtering operator $\hat{\mathcal{Q}}$, the resulting spectra, and correlation functions of the synthetic fields used in this work are discussed in a later section.

  \begin{figure}[!t]
    \centering
    \includegraphics[width=1.0\linewidth]{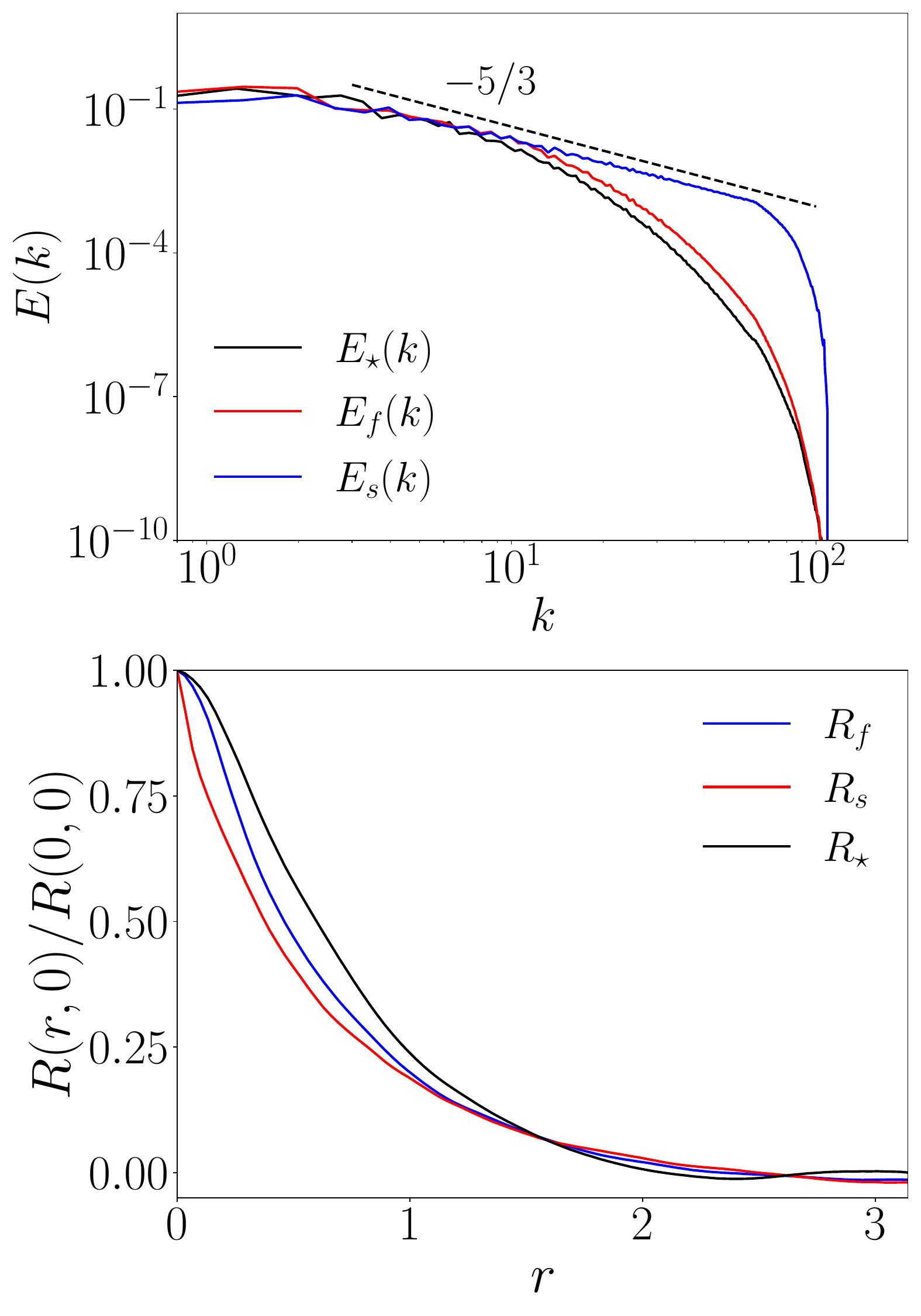}
\put(-220,330){(a)}
    \put(-220,170){(b)}
    \caption{(a) Kinetic energy spectra with an additional arbitrary straight line of slope $-5/3$. (b) Normalized radial correlation function.}
    \label{fig: Velocity Energy Spectrum}
\end{figure}

 \subsection{Parameters for Generating Synthetic Velocity Fields}
To simulate mixing by synthetically generated stochastic velocity field, we use the same pseudospectral solver used in 3D DNS, to solve the modified density equation \eqref{eq: density} with a stochastically generated velocity field. To obtain the desired spectrum, we use the filtering kernel  
\begin{equation}
    \hat{\mathcal{Q}} = \mathcal{A} (1 + \lambda^2 k^2)^{-n},
    \label{eq: Karman-Obukhov's spectrum}
\end{equation}
where $\mathcal{A}=12$ is a constant to scale the energy spectrum, and  $\lambda$ denotes a length scale. We also choose the initial condition as 
\begin{equation}
    \hat{\boldsymbol{\eta}} (\boldsymbol{k},0) = \left(\frac{\varepsilon}{\tau}\right)^{1/2}\hat{\mathcal{Q}}  \hat{\boldsymbol{\chi}},
\end{equation}
to obtain a homogeneous isotropic velocity field from the beginning. The exponent $n$ in  \eqref{eq: Karman-Obukhov's spectrum} can be derived using the desired scaling of the kinetic energy spectrum. For instance, to obtain $-5/3$ scaling for large $\lambda k$ values, 
\begin{equation}
    E_s(k) \approx k^{-5/3} \approx k^4(\lambda^2k^2)^{-2n},
\end{equation}
we obtain $n=17/12$, choosing $\lambda=1$. Choosing $\tau=T_L$, we obtain the energy spectra $E_s(k)$ shown in figure \ref{fig: Velocity Energy Spectrum}a. We see that spatial filtering creates three distinct energy regimes - the energy-containing scales, the inertial subrange and the dissipation range, a standard feature of the so-called Karman-Obhukov spectra~\cite{careta1993stochastic}. The resulting energy spectra and the radial correlation function $R_s$ (using \eqref{eq: R_Ek_relation}) are shown in figures~\ref{fig: Velocity Energy Spectrum}a and b respectively, compared against the energy spectrum and radial correlation obtained from the incompressible 3D DNS discussed in previous section. Figure~\ref{fig: Velocity Energy Spectrum} also shows that the energy spectrum obtained from the synthetic cases has a wider inertial range when compared to the incompressible 3D DNS used for initialising the compressible simulations. To isolate the effect on mixing (if any), we modify the synthetic velocity field through the spatial filter,
\begin{equation}
    \boldsymbol{\hat{u}}_f(k) = \boldsymbol{\hat{u}}_s(k)  \exp \left[-5.5 \left(\frac{k - 10}{64\sqrt{3}-10}\right)^{1.05}\right ].
    \label{eq: spatial filter}
\end{equation}
The corresponding $E_f(k)$ and $R_f(r)$ are also shown in figures~\ref{fig: Velocity Energy Spectrum}a and b, respectively. 

\begin{table*}[!t]
\centering
\begin{tabular}{|l|c|c|c|}
\hline
\multirow{2}{*}{} & \multicolumn{1}{c|}{\textbf{Diffusion}} & \multicolumn{2}{c|}{\textbf{Synthetic}} \\ \hline
                  & {Different $\overline{\rho}$} & ${E(k) = E_f(k)}$ & ${E(k) = E_s(k)}$ \\ \hline
${At = 0.75}$       & 1a ($\overline{\rho}$ = 6.43)   & 1c & 1e \\ \hline
${At = -0.75}$      & 2a ($\overline{\rho}$ = 0.18)  & 2c & 2e \\ \hline
{Passive}           & 3a ($\overline{\rho}$ = 1.00) & 3c & 3e \\ \hline
\end{tabular}
\caption{Parameter space for simulations and their indicators.}
\label{tab: Synthetic-cases}
\end{table*}
 To understand the effect of density gradients on the species diffusion alone, we run one case for each of the positive and negative Atwood numbers without any stirring and compare them with a passive scalar case. These cases are labeled using ``a'' in table \ref{tab: Synthetic-cases}. We simulate six cases using the synthetically generated stochastic field. In three of those cases, we use $E(k) = E_f(k)$. These are labeled with ``c'' in table \ref{tab: Synthetic-cases}. In the other three cases, we keep the spectra with $-5/3$ scaling in the inertial subrange, labeled using ``e'' in table~\ref{tab: Synthetic-cases}.

 \subsection{Stirring Enhanced Diffusion}
\label{sec: Stirring Enhanced Diffusion}
The action of stirring by a stochastic incompressible flow field $\boldsymbol{u}_c$ is reflected in \eqref{eq: density} through the advection term. Since the process of homogenization is a non-equilibrium inhomogeneous process, any statistical analysis can be done only through ensemble averaging, without any replacement from time averaging (using ergodicity) or spatial averaging. To this end, we follow \citeauthor{klyatskin1994statistical}'s
study on the diffusion of a passive tracer \cite{klyatskin1994statistical} to show how the synthetic stochastic fields enhance mixing. Let $\boldsymbol{u}_c$ be a stochastic field in space and time. Taking ensemble average in \eqref{eq: density} we obtain,
\begin{equation}
    \frac{\partial \avg{\rho}}{\partial t} + \avg{\boldsymbol{u}_c\cdot\nabla\rho} = \frac{1}{\mathrm{Re}\mathrm{Sc}}\nabla^2\avg{\rho}.
    \label{eq: density_ensemble_avg}
\end{equation}
The correlation term $\avg{\nabla\cdot\left(\boldsymbol{u}_c\rho\right)}$ in \eqref{eq: density_ensemble_avg} captures the average effect of the random velocity field $\boldsymbol{u}_c$ on mixing. This term can be split using the variational derivatives~\cite{klyatskin1996short, novikov1965functionals} for a zero-mean Gaussian $\boldsymbol{u}_c$ field yielding, 
\begin{align}
 &\frac{\partial \avg{\rho}}{\partial t}   = \frac{1}{\mathrm{Re}\mathrm{Sc}}\nabla^2\avg{\rho}-\nonumber\\
 &\int\displaylimits^t_0dt'\int\displaylimits_B d\boldsymbol{y}R_{ij}(\boldsymbol{x},t;\boldsymbol{y},t')\frac{\partial}{\partial x_i}\avg{\frac{\delta \rho(\boldsymbol{x},t)}{\delta u_{cj}(\boldsymbol{y},t')}},
 \label{eq: Novikov-Furutsu}
\end{align}
where the final term on the RHS is obtained using the variational derivatives (defined using $\delta$), assuming $\boldsymbol{u}_c$ is Gaussian in time and space~\cite{klyatskin1996short, novikov1965functionals} and $R_{ij}$ is defined as
\begin{equation}
    R_{ij}(\boldsymbol{x},t,\boldsymbol{y},t') = \avg{{u}_{ic}(\boldsymbol{x},t)u_{jc}(\boldsymbol{y},t')}.
\end{equation}
If $\boldsymbol{u}_c$ is a delta-correlated (in time) zero-mean stochastic incompressible field with $R_{ij}(\boldsymbol{x},t,\boldsymbol{y},t') = \tilde{R}_{ij}(\boldsymbol{x},t,\boldsymbol{y})\delta(t-t')$, then \eqref{eq: Novikov-Furutsu} reduces to~\cite{klyatskin1994statistical},
\begin{equation}
    \frac{\partial \avg{\rho}}{\partial t} = \left(\tilde{R}_{ij}(\boldsymbol{x},\boldsymbol{x},t)\frac{\partial^2}{\partial x_i\partial x_j} + \frac{1}{\mathrm{Re}\mathrm{Sc}}\nabla^2\right)\avg{\rho}.
    \label{eq: delta_correlated_mixing}
\end{equation}
Additionally, for an incompressible $\boldsymbol{u}_c$ with some correlation time $\tau$, a set of coupled equations in $\avg{\rho}$ and $\avg{\delta \rho/\delta \boldsymbol{u}_c}$ can be derived~\cite{lipscombe1991convection, klyatskin1994statistical}. Equation~\eqref{eq: delta_correlated_mixing} shows that the velocity correlations supply an additional diffusion coefficient. It is also interesting to note that for anisotropic correlations, such diffusion is also anisotropic. Since turbulence is known to be non-Gaussian~\cite{wilczek2011velocity}, simplifying \eqref{eq: density_ensemble_avg} for an actual turbulent field is difficult. In this work, both the synthetic fields and turbulent velocity fields are not delta-correlated. Hence \eqref{eq: Novikov-Furutsu} can only be further simplified using approximations~\cite{lipscombe1991convection}. Furthermore, since the mixing process is non-equilibrium in nature, an ensemble average of several DNS simulations would have to be computed, which may approach the behavior governed by \eqref{eq: Novikov-Furutsu}. 

In sections below, we analyze the mixing dynamics due to both the synthetic velocity field as well the turbulent velocity field in 3D DNS. We show that although synthetic fields show similar mixing trends to DNS turbulent fields, they are unsuitable for analyzing the transient evolution quantitatively. Though the above equations are derived for density, homogenization of density inhomogeneities will also result in the homogenization of concentration inhomogeneities. Therefore, we shall use concentration fields to track homogenization.

\section{Results}
\label{sec: Results}
\subsection{Effect of density gradients on diffusion}
To highlight the difference between diffusion of active scalars and passive scalars in 3D, we consider cases 1-3a in table~\ref{tab: Synthetic-cases}. We use $\boldsymbol{u}_c=0$ in~\eqref{eq: density} and \eqref{eq: species_gov_equation}. For passive scalars, the concentration field simply diffuses. However, as also discussed by~\citeauthor{jossy2025mixing}~\cite{jossy2025mixing} for two dimensions, the active scalar diffusion of concentration fields consists of density weighted diffusion flux, which results in a nonlinear diffusion term. This nonlinear term results in an apparent expansion and contraction of the initially inhomogeneous spherical blob. 
Without any stirring, the density and scalar concentration equations reduce to,
\begin{align}
    &\frac{\partial \rho}{ \partial t} = \frac{1}{\mathrm{Re}  \mathrm{Sc}}
    \nabla^{2}\rho,~~\rho(Y_c) = \frac{\chi\rho_c}{1 + Y_c\left(\chi- 1 \right)},
    \label{eq: pure diffusion density}\\
    & \frac{\partial Y_{c}}{\partial t}  = \frac{1}{  \mathrm{Re} \mathrm{ Sc} }\left( \frac{2\nabla \rho \cdot \nabla Y_{c}}{\rho} + \nabla^{2}Y_{c} \right).
     \label{eq: species equation}
\end{align}
\begin{figure}[!t]
    \centering
\includegraphics[width=1.0\linewidth]{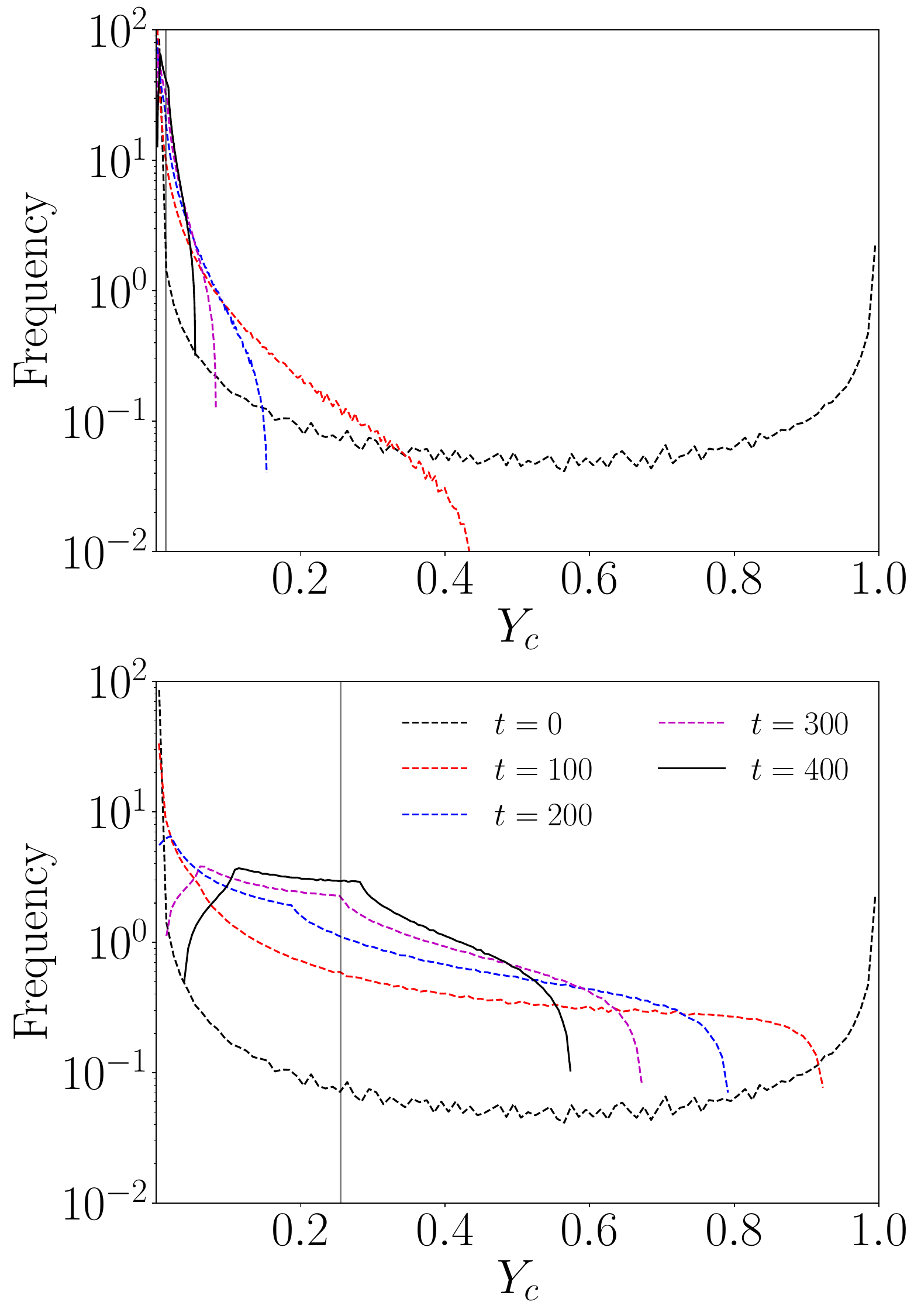}
\put(-220,330){(a)}
    \put(-220,160){(b)}
    \caption{(a) Normalized spatial distribution of the spherical species at different time instances for pure diffusion for case 1a and (b) case 2a. The vertical lines indicate the final mixed-state (equilibrium concentration) value. The equilibrium concentration for case 1a is 0.0142, and for  2a is 0.255.}
    \label{fig: PDF Species  Pure diffusion concentration time series}
\end{figure}
The above equations show that $\rho(Y_c)$ represents a Cole-Hopf transformation for the nonlinear concentration field equation. Additionally,~\eqref{eq: species equation} shows that the nonlinear term results in the movement of the boundary between the spherical and the surrounding species with a velocity proportional to $-\nabla\rho/\rho$. Since, for $At>0$, $\nabla\rho > 0$ at the boundary (radially outward), the boundary moves inward, and vice-versa. An analytical representation of the solution of~\eqref{eq: species equation} can be derived using the unsteady heat equation solution~\cite{evans2022partial}. 
As an inhomogeneous mixture evolves, the concentration homogenization results in the spatial distribution of the concentration field approaching a Dirac delta function. Figures~\ref{fig: PDF Species  Pure diffusion concentration time series}a and b show the spatial distribution frequency of $Y_c$ for cases 1a and 2a. While complete homogenization due to pure diffusion is approached at $t\to\infty$ for both the cases, a lighter spherical inhomogeneity ($At>0$) seemingly approaches this homogenization faster compared to a heavier inhomogeneity. 
\begin{figure}[!t]
    \centering
    \includegraphics[width=1.0\linewidth]{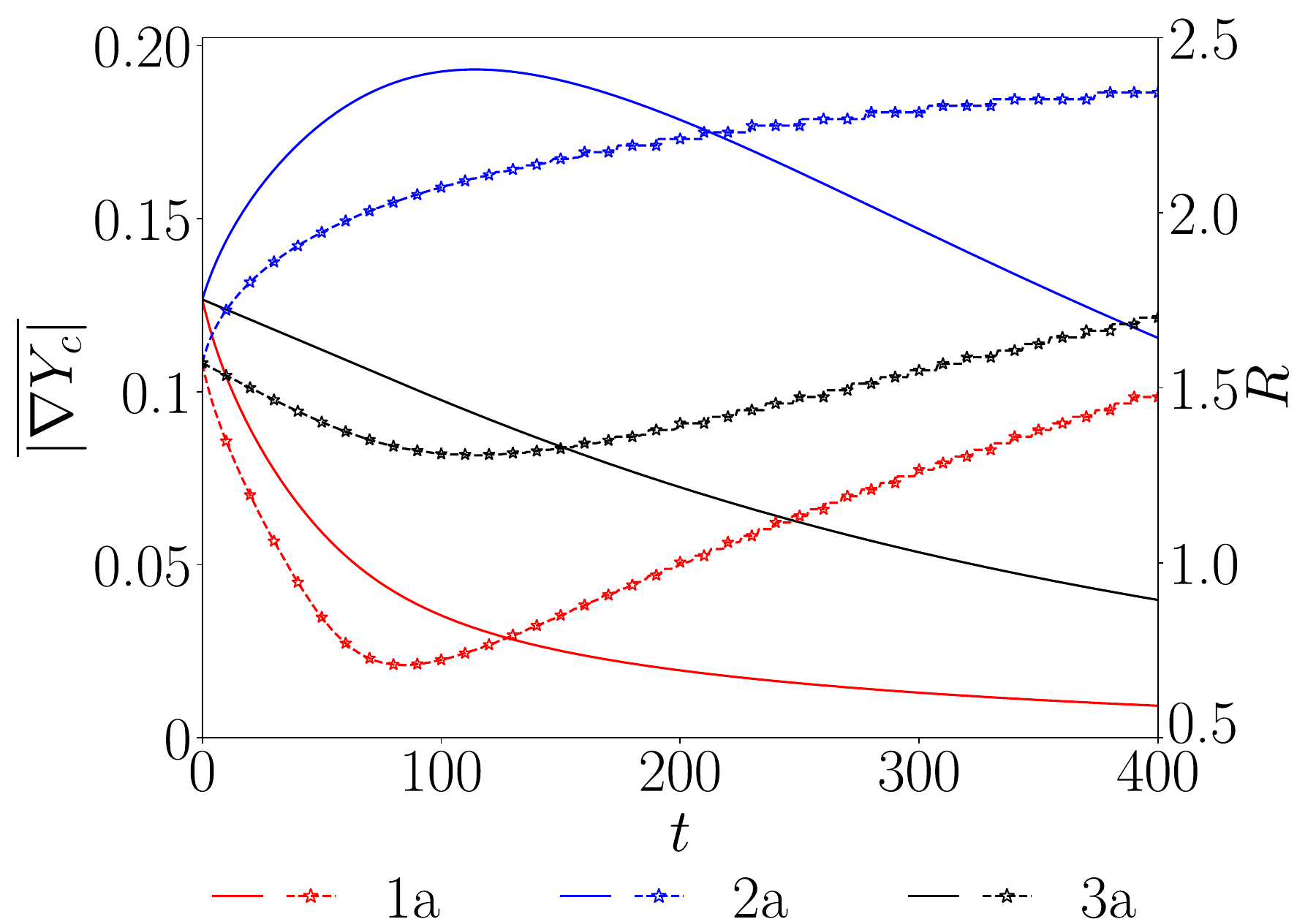}
    \caption{Evolution of volume-averaged concentration gradients under pure diffusion (denoted by solid lines) and the location of the maximum value of the concentration gradient (denoted by dashed lines with markers). Note that the colour schemes remain the same for both quantities. }
    \label{fig: Pure Diffusion Grad Yc}
\end{figure}
Figure~\ref{fig: Pure Diffusion Grad Yc} shows the evolution of spatially averaged magnitude of concentration gradient $\overline{|\nabla Y_c|}$ for 1a, 2a, and 3a cases. As a mixture evolves towards homogenization, $\overline{|\nabla Y_c|}\to 0$. For a passive scalar diffusion (3a), pure diffusion results a monotonically decreasing $\overline{|\nabla Y_c|}$. In active scalars, $\overline{|\nabla Y_c|}$ decays faster than a passive scalar for $At>0$. Contrastingly, for $At<0$, $\overline{|\nabla Y_c|}$ increases before decreasing. The faster decay of $\overline{|\nabla Y_c|}$ for $At>0$ is due to the inward movement of the boundary of the inhomogeneity, which results in a decrease in the volume over which $\nabla Y_c$ exists. Similarly, the apparent increase in $\overline{|\nabla Y_c|}$ for $At<0$ in the initial stages is due to the outward movement of the boundary. The Laplacian term in \eqref{eq: species equation} overcomes the boundary movement effect and eventually, $\overline{|\nabla Y_c|}$ approaches $0$ as the inhomogeneities diffuse. 
This fundamental difference between an active scalar and a passive scalar diffusion results in different mixing dynamics in a stochastic turbulent field, as discussed in the next section.

\subsection{Comparison of Active scalars and Passive scalars }
\begin{figure*}[!t]
    \centering
\includegraphics[width=1.0\linewidth]{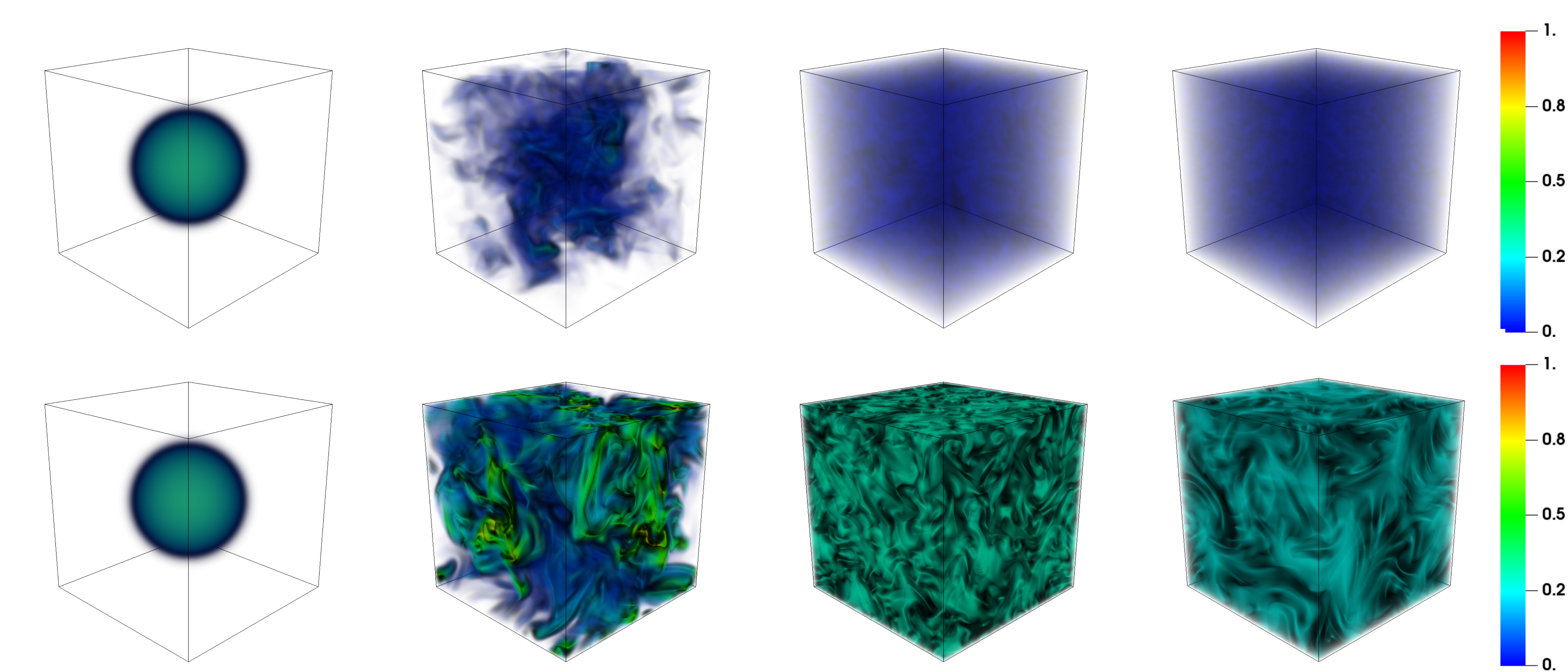}
\put(-470,190){(a)}
\put(-350,190){(b)}
\put(-250,190){(c)}
\put(-140,190){(d)}
\put(-430,200){$t=0$}
\put(-320,200){$t=2.5$}
\put(-200,200){$t=10$}
\put(-90,200){$t=25$}
\put(-470,90){(e)}
\put(-350,90){(f)}
\put(-250,90){(g)}
\put(-140,90){(h)}
\put(-430,95){$t=0$}
\put(-320,95){$t=2.5$}
\put(-200,95){$t=10$}
\put(-90,95){$t=25$}
    \caption{Concentration evolution of the spherical blob species of (a-d) $At =0.75$ (case 1d) and (e-h) $At =- 0.75$ (case 2d)  at different times.}
    \label{fig: 3D figures of  Yc with stirring}
\end{figure*}
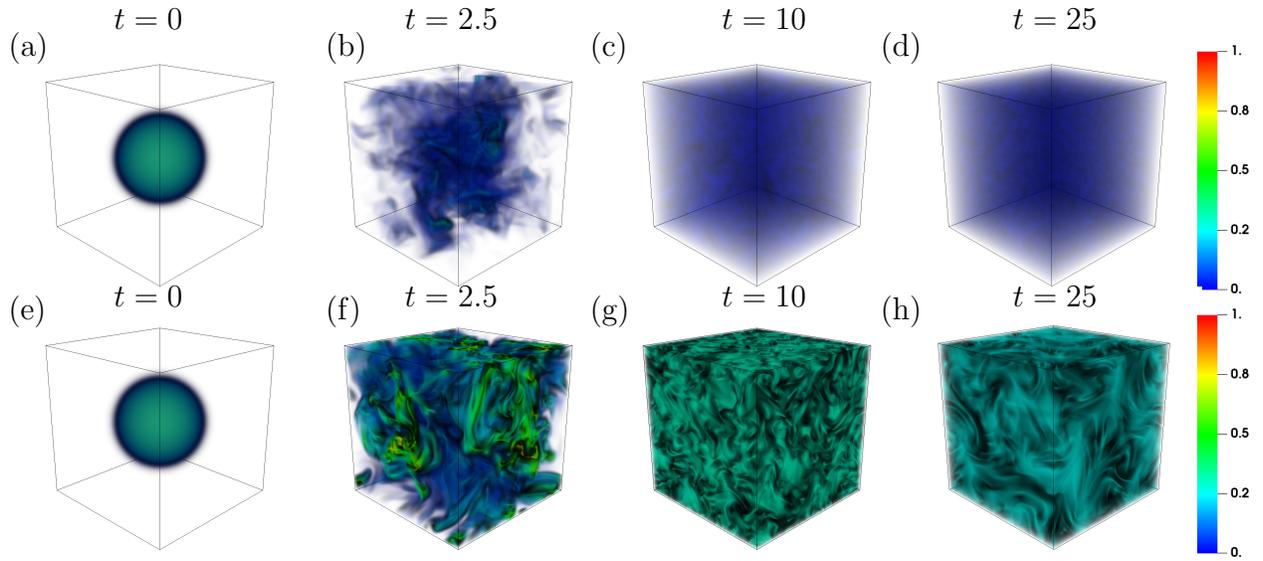
\begin{figure*}[!t]
    \centering
\includegraphics[width=1.0\linewidth]{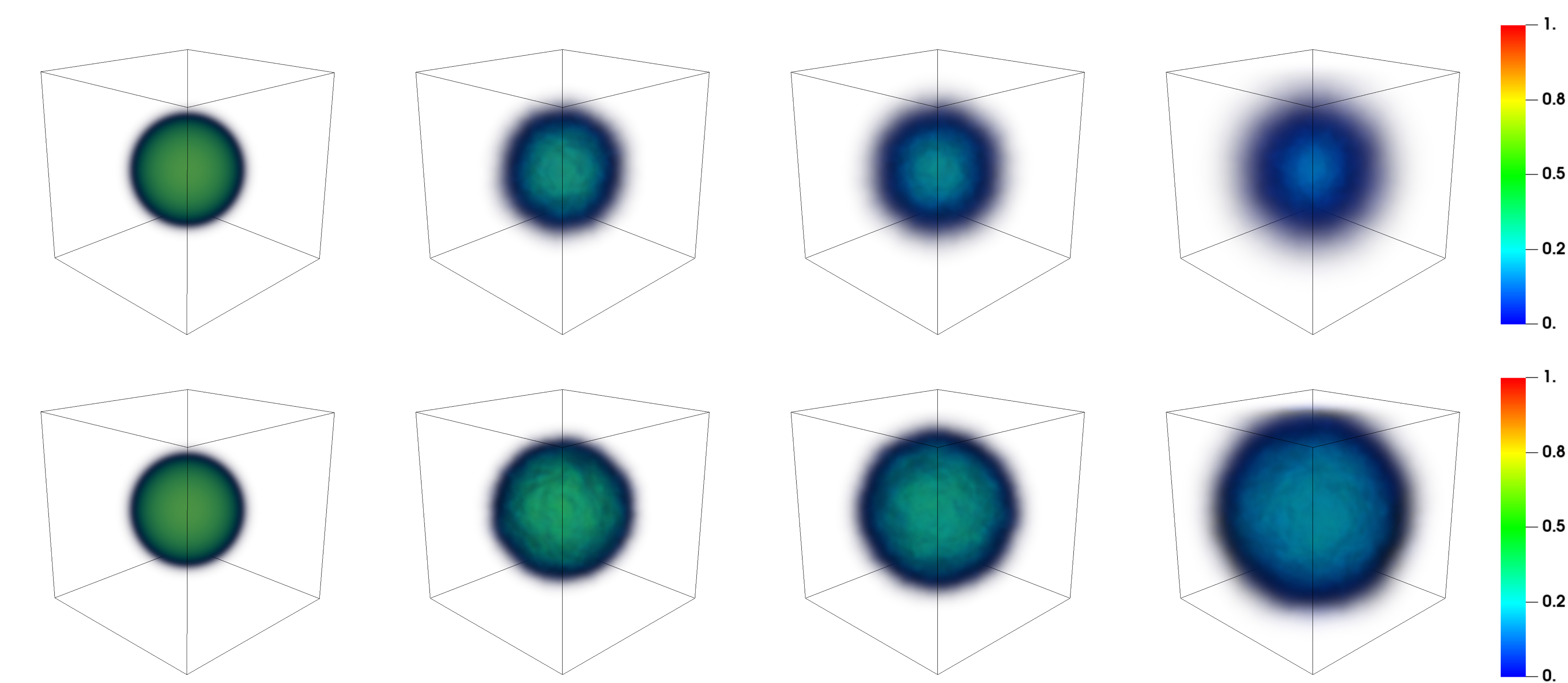}
\put(-470,190){(a)}
\put(-350,190){(b)}
\put(-250,190){(c)}
\put(-140,190){(d)}
\put(-430,200){$t=0$}
\put(-320,200){$t=2.5$}
\put(-200,200){$t=10$}
\put(-90,200){$t=25$}
\put(-470,90){(e)}
\put(-350,90){(f)}
\put(-250,90){(g)}
\put(-140,90){(h)}
\put(-430,95){$t=0$}
\put(-320,95){$t=2.5$}
\put(-200,95){$t=10$}
\put(-90,95){$t=25$}
    \caption{Concentration evolution of the spherical blob species
    of (a-d) $At =0.75$ (case 1c) and (e-h) $At=-0.75$ (case 2c) at different times. }
    \label{fig: 3D figures of  Yc with stirring synthetic}
\end{figure*}

Stirring enhances the diffusion process in two ways. Stirring increases the surface area across which the molecular diffusion occurs, allowing for more homogenization by diffusion. Stirring also increases the gradients, thereby increasing the rate of diffusion. Figures \ref{fig: 3D figures of  Yc with stirring}(a-d) and figures \ref{fig: 3D figures of  Yc with stirring}(e-h) show the evolution of the blob for $At=0.75$ and $At= - 0.75$ respectively at different times for the 3D DNS cases. Figures \ref{fig: 3D figures of  Yc with stirring synthetic}(a-d) and figures \ref{fig: 3D figures of  Yc with stirring synthetic}(e-h) show the evolution of the blob for $At=0.75$ and $At= - 0.75$ respectively at different times with synthetic stirring. We see that the boundaries are stretched and corrugated by the stirring process. The stretching is more pronounced for the DNS cases, while the synthetic turbulence cases produce just small scale corrugations on the surface. Figures \ref{fig: 3D figures of  Yc with stirring}b and \ref{fig: 3D figures of  Yc with stirring}f show that stirring folds and stretches the interface in both positive and negative Atwood number cases. We see that at later times (figure \ref{fig: 3D figures of  Yc with stirring}d and figure \ref{fig: 3D figures of  Yc with stirring}h), the positive Atwood number case is seen to be more homogenized that the negative Atwood number case.

\subsubsection{Concentration Gradients}
\begin{figure}[!t]
    \centering
\includegraphics[width=1.0\linewidth]{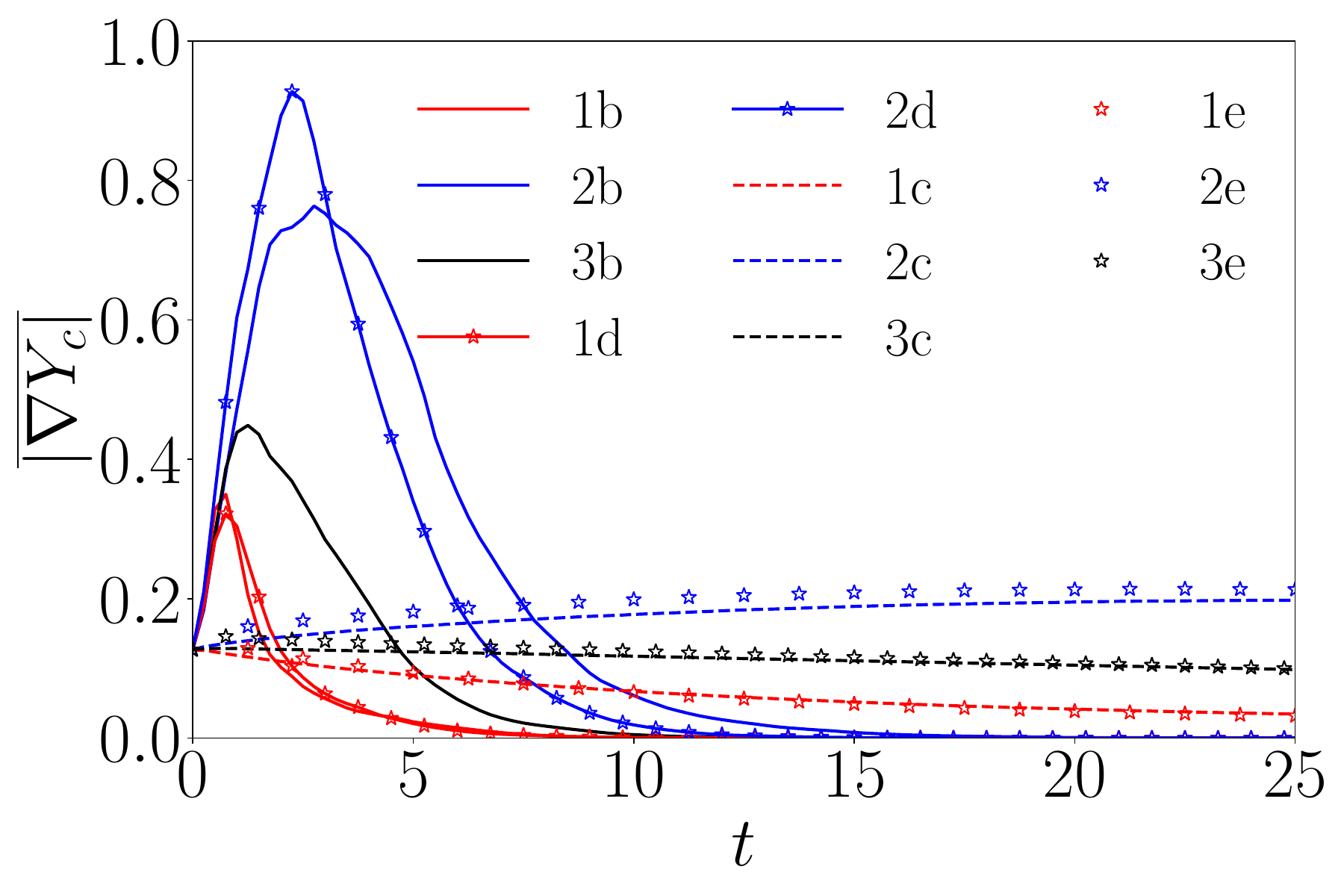}
\caption{The evolution of volume-averaged concentration gradients. For DNS and synthetic cases, the lighter mixtures have a higher increase in the volume-averaged concentration gradients. But this higher increase does not indicate faster mixing than denser mixtures.}
    \label{fig: Grad Yc with stirring}
\end{figure}

\begin{figure*}[!t]
    \centering
\includegraphics[width=1.0\linewidth]{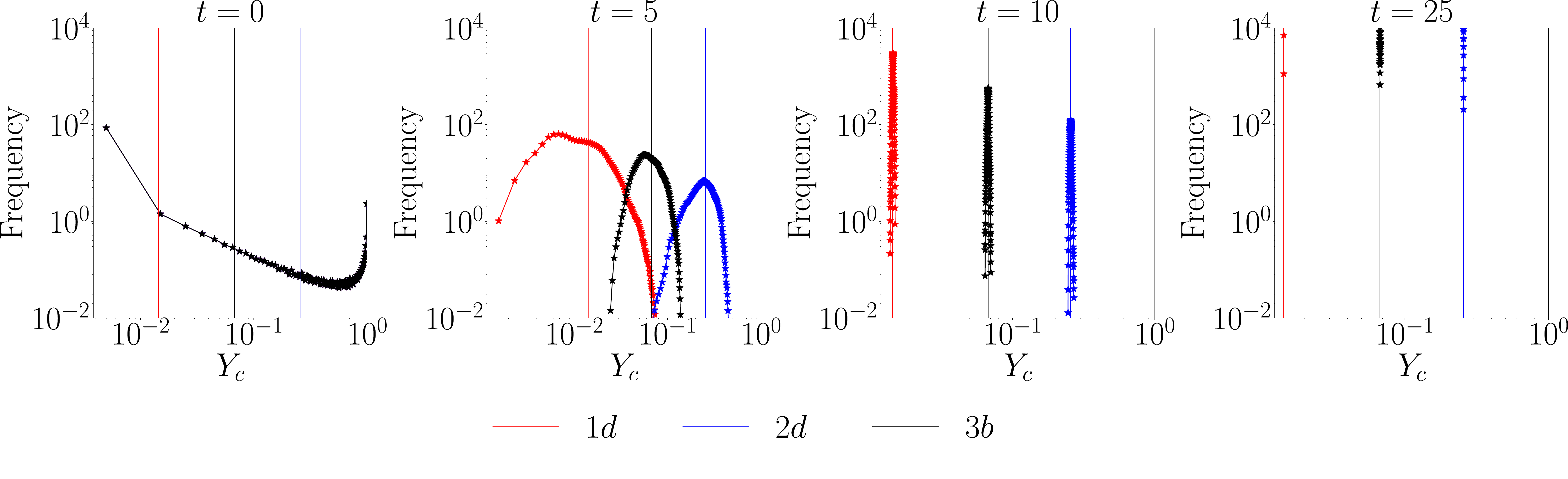}
\put(-470,140){(a)}
\put(-355,140){(b)}
\put(-237,140){(c)}
\put(-121,140){(d)}
    \caption{ Normalized spatial distribution of concentration at different time instances. The vertical lines indicate the final mixed-state value. Positive Atwood number cases near the final homogenized state faster than negative Atwood number cases, even though the negative Atwood number spreads more. The equilibrium concentration for case 1b is 0.0142, for 2b is 0.255, and for 3b is 0.0673.}
    \label{fig: PDF Species  concentration}
\end{figure*}

To quantify the effect of stirring, we calculate the volume-averaged concentration gradient magnitude $\overline{|\nabla Y_c|}$ similar to the pure diffusion cases. Figure \ref{fig: Grad Yc with stirring} shows the evolution of $\overline{|\nabla Y_c|}$ for all the cases with stirring in Table \ref{tab: Synthetic-cases}. In all the cases, we see that stirring increases the concentration gradients as compared to the pure diffusion cases in figure~\ref{fig: Pure Diffusion Grad Yc}. For both the DNS and synthetic field cases, the concentration gradients increase more than the passive scalar case for the negative Atwood number cases. However, the increase is less for the positive Atwood number when compared to the passive scalar. The differences in the increase of spatially concentration gradient magnitude are due to the combined effect of density-driven nonlinear diffusion term (discussed in the previous section) and stirring. The outward expansion in the negative Atwood number cases increases the surface area. It causes the gradients to exist over a larger area. Similarly, the inward contraction of the interface decreases the surface area across which gradients exist for the positive Atwood number cases. Stirring increases the rate at which the trends seen in the pure diffusion cases occur such that the denser mixture ($At =0.75$) homogenizes faster than the lighter mixture ($At = -0.75$).
The drop in volume-averaged concentration gradients is an indication of homogenization. However, the expansion and contraction effects of the interface affect the volume averaging. Figures~\ref{fig: PDF Species  concentration}(a-d) show the normalized histogram of the spatial distribution of $Y_c$ for cases 2a, 2b, and 2c at different times. As the mixing evolves, the denser mixture ($At=0.75$) distribution is centered about its final mixed state value while the lighter mixture $At=-0.75$ has a skewed distribution. We see in figure \ref{fig: PDF Species  concentration}c that even though the average concentration gradient magnitude $\overline{|\nabla Y_c|}$ approaches zero for the lighter mixture ($At=-0.75$), the distribution is still spread out compared to that for the denser mixture ($At=0.75$). This indicates that low value of $\overline{|\nabla Y_c|}$ is not a true measure of homogenization. Figure~\ref{fig: PDF Species  concentration}d shows the scenario when a mixture is perfectly homogenized mixture and the spatial distribution follows a Dirac delta function centered about the value at the final homogenized state. 

The trend of higher concentration gradient magnitude $\overline{|\nabla Y_c|}$ for lighter mixtures ($At<0$) is followed irrespective of the stirring method. However, we see that the rate of increase and the rate of decrease of $\overline{|\nabla Y_c|}$ are larger for the DNS than the synthetic cases. This indicates that the synthetic velocity fields produce weaker stirring effects.

\subsubsection{Mixing Quality}

 \begin{figure}[!b]
    \centering
    \includegraphics[width=1.0\linewidth]{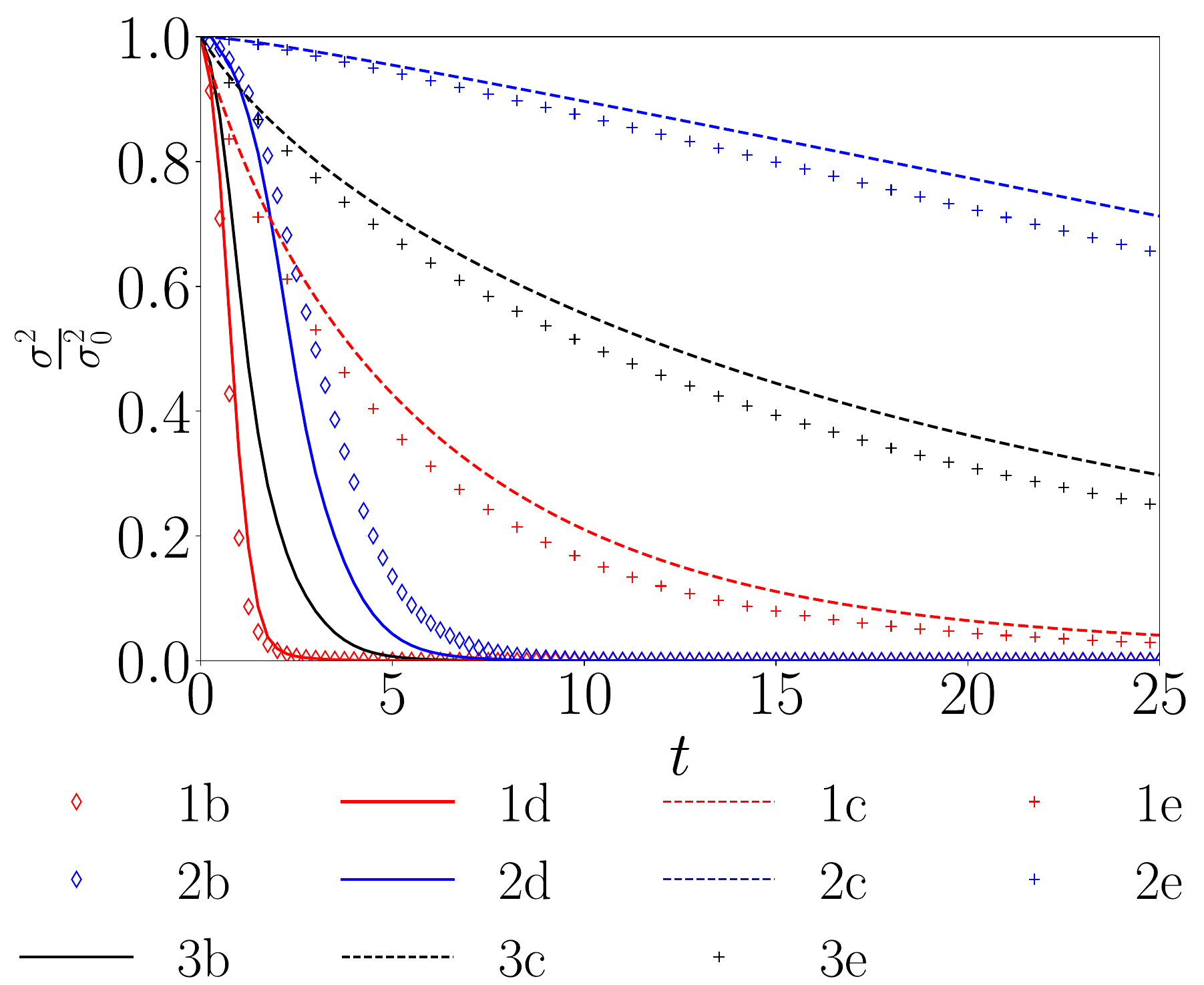}
    \caption{Evolution of Mixing quality normalized by the initial value. The DNS cases  stir the mixtures more efficiently when compared to the synthetic cases. }
    \label{fig: Evolution of Mixing quality}
\end{figure}

Volume-averaged concentration gradient magnitude $\overline{|\nabla Y_c|}$ is not a suitable mixing metric for active scalar mixing. Figure \ref{fig: Grad Yc with stirring} shows that the positive Atwood number cases homogenize faster than the negative Atwood number cases; however, the volume averaging reduces its reliability since the volume over which $\nabla Y_c$ exists changes due to both turbulent mixing and nonlinear diffusion. To evaluate the mixing measure more appropriately, we calculate the mixing variance defined by 
\begin{equation}
    \sigma^{2}(t) = \overline{( Y_{c} - Y^{eq}_{c})^{2}},
\end{equation}
 where $Y^{eq}_{c}$ is the concentration at the homogenized state. Mixing variance tends to zero for a fully homogenized case. Figure \ref{fig: Evolution of Mixing quality} shows the evolution of $\sigma^{2}(t)$ normalized by initial value for all the cases. As indicated by figure \ref{fig: Grad Yc with stirring}, a lighter blob (denser mixture, $At > 0$) homogenizes faster when compared to a heavier blob (lighter mixture, $At < 0$), with the passive scalar case being in between them. \Revision{}{It is interesting to note that though the negative Atwood number cases have higher concentration gradients and also fill the volume faster (compare DNS cases, figure \ref{fig: 3D figures of  Yc with stirring}b with the synthetic cases, figure \ref{fig: 3D figures of  Yc with stirring}f and also figure~\ref{fig: 3D figures of  Yc with stirring synthetic}d with figure \ref{fig: 3D figures of  Yc with stirring synthetic}h), it does not result in better mixing. Unlike passive scalar, whose diffusion phase begins once the entire volume is filled \cite{orsi2021scalar}, the presence of density gradients alters the time at which the diffusion phase starts. The above results indicate that in the presence of density gradients, the spread of the constituent species to all parts of the volume need not necessarily indicate better mixing} .
\subsubsection{Effect of density gradients on velocity}
\label{sec: Effect of density gradients on velocity}
\begin{figure}[!t]
    \centering
    \includegraphics[width=1.0\linewidth]{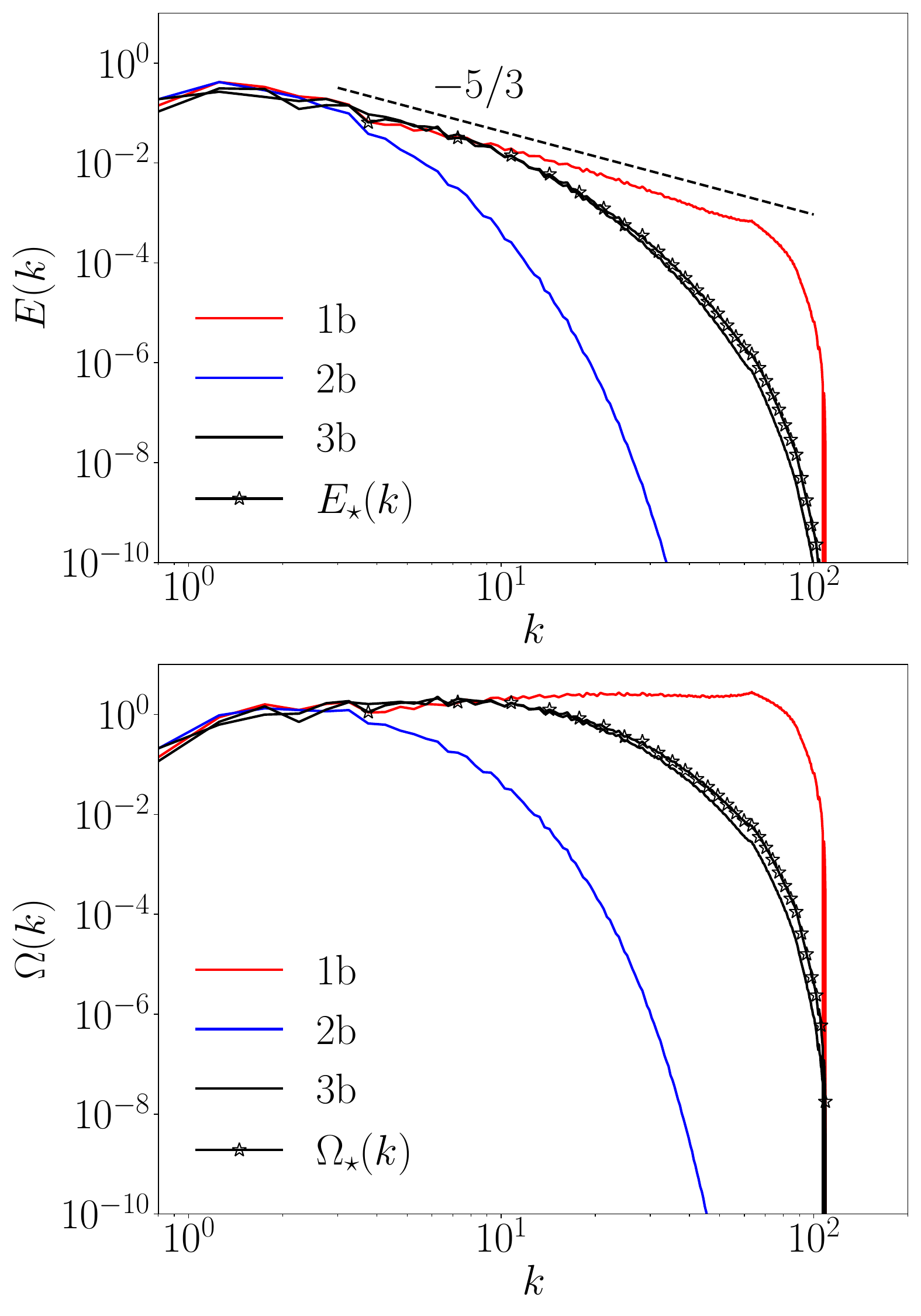}
        \put(-200,330){(a)}
    \put(-200,170){(b)}
    \caption{(a) Kinetic energy spectra and (b) enstrophy spectra for the DNS cases with different mean mixture density. Both the spectra are broader than the incompressible spectra for the denser mixture and steeper for the lighter mixture. Passive scalars exhibit identical spectra as the incompressible initial condition.}
    \label{fig: combined Spectra from DNS with different initial mean density}
\end{figure}

The primary contrast between active and passive scalar mixing is based on the constituents' ability to affect the mixing process, which we refer to as the two-way coupling. The previous section shows that the combined effect of density gradients and stirring affects the homogenization process of active scalars. In this section, we investigate the effect of density gradients on the stirring process (the advection term). In synthetic field cases, the velocity is not affected by inhomogeneous states as it is generated at each time instant. Though the velocity field is synthetically generated, the enforcement of~\eqref{eq: velocity constraint} accounts for some features of the two-way coupling.  As discussed earlier, the action of stirring is reflected through velocity. In the momentum equation \eqref{eq:momentum_equation}, the density affects the pressure gradient term and the diffusion terms. These are the terms through which density can affect the velocity field responsible for stirring. Misalignment of pressure and density gradients are responsible for generating vorticity by baroclinity. Similarly, misalignment of bulk viscous stresses and density also generates vorticity. Shear viscous stresses are responsible for the diffusion of vorticity. Density differences may enhance or weaken the effect of vorticity diffusion, depending on the relative gradients. 

Given that the variable density affects different vorticity generation and diffusion mechanisms, we expect a generation or destruction of scales of the turbulence based on density gradients. Figure \ref{fig: combined Spectra from DNS with different initial mean density} shows the kinetic energy $E(k)$ and the enstrophy $\Omega(k) = |\hat{\boldsymbol{\omega}}|^2/2$ spectra of the DNS cases with different mean mixture density values. For the passive scalar case, we see the spectra remain same as the initial incompressible spectra, indicating no coupling. However, the denser mixture has wider spectra than the initial incompressible spectra, while the lighter mixture's spectra become more steep. This occurs since we keep the forcing $\boldsymbol{F}$ parameters identical in \eqref{eq:momentum_equation}, even though the average density $\overline{\rho}$ changes drastically for lighter and heavier mixtures. This is primarily due to our choice of keeping $\rho_c=1$ in the initial conditions. For same statistical forcing $\boldsymbol{F}$, the viscous dissipation is pronounced for lighter mixtures ($At<0$) and reduced for denser mixtures ($At>0$). This results in a different range of scales generated in the two mixtures. 
\begin{figure}[!t]
    \centering
\includegraphics[width=1.0\linewidth]{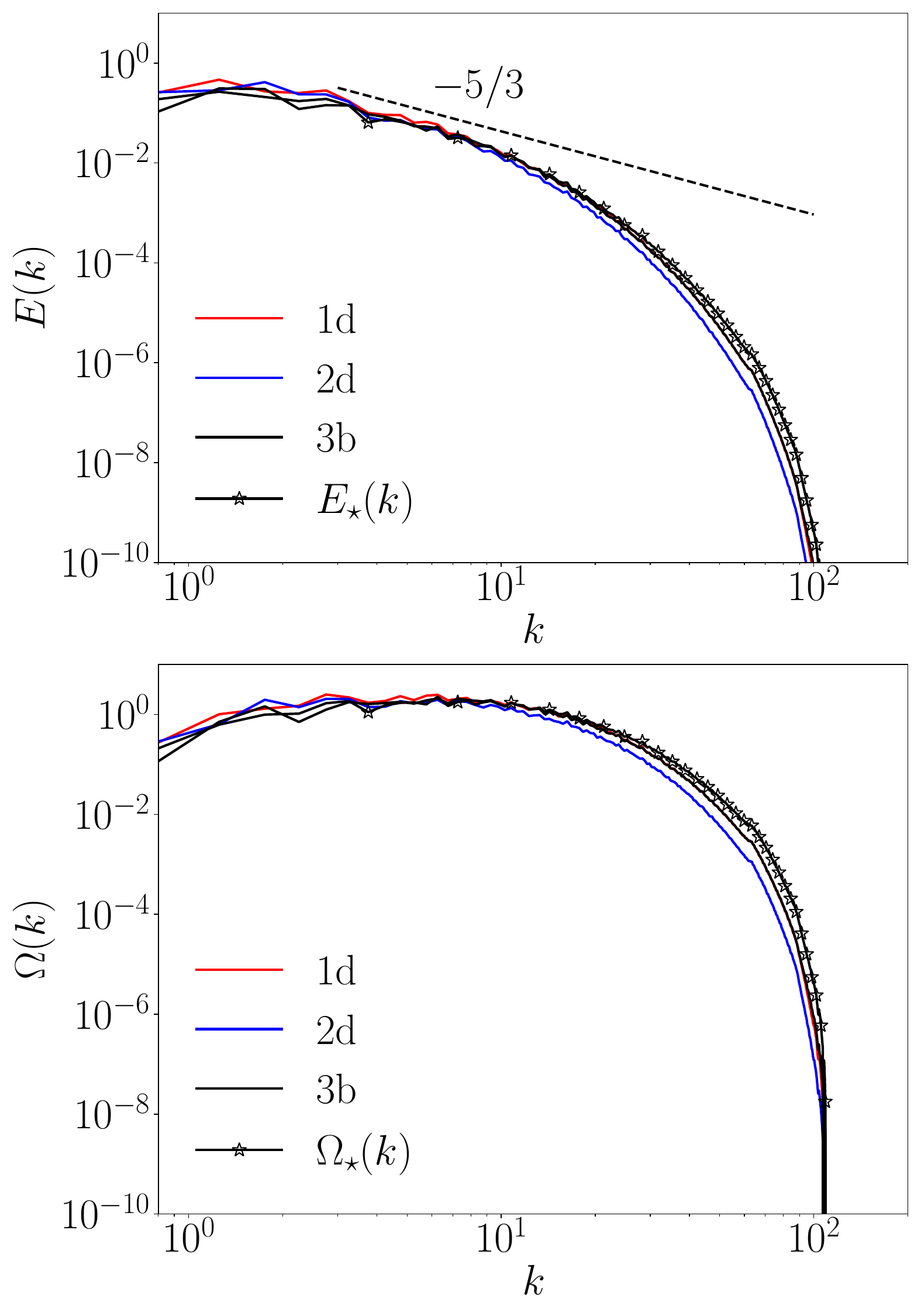}
\put(-200,330){(a)}
    \put(-200,170){(b)}
    \caption{(a) Kinetic energy spectra and (b) enstrophy spectra for the DNS cases with same spatially averaged mixture density. Both the spectra are the same for all cases, indicating a negligible two-way coupling.}
    \label{fig: combined Spectra from DNS with same initial mean density}
\end{figure}
To confirm this, we use cases (1d and 2d) where we initialize the average mixture density $\overline{\rho}$ to be identical, but achieve different density ratios. Figure \ref{fig: combined Spectra from DNS with same initial mean density} shows that the kinetic energy and the enstrophy spectra are similar, indicating that there is a negligible effect of density gradients on the stirring velocity in the current study. Note that no changes in trends of homogenization behavior are observed, irrespective of the values of $\overline{\rho}$ (c.f. figure \ref{fig: Grad Yc with stirring}). The current study shows that for low Mach turbulence, negligible two-way coupling is observed in the spectra. Effects on velocity correlations in time at different length scales are deferred to future studies. \Revision{}{For high Mach turbulence, shocklets may develop. \citeauthor{jossy2025mixing} \cite{jossy2025mixing} have shown that weak shocks ($M\approx 1.1$) generate vorticity due to very strong baroclinic torque, which  plays a  major role in the mixing process in 2D. This baroclinic vorticity falls as the mixture homogenizes.}  Previous studies of variable density mixing by incompressible turbulence use specially derived pressure equations~\cite{sandoval1995dynamics,livescu2020turbulence,livescu2008variable}. The current study also shows that using weakly compressible equations is sufficient to study incompressible mixing without acoustic wave turbulence interference.   
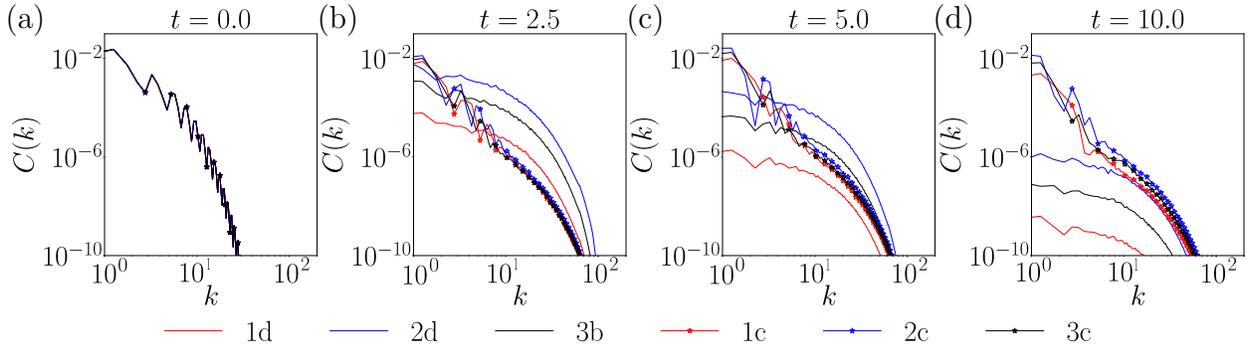
\begin{figure*}[!t]
    \centering
\includegraphics[width=1.0\linewidth]{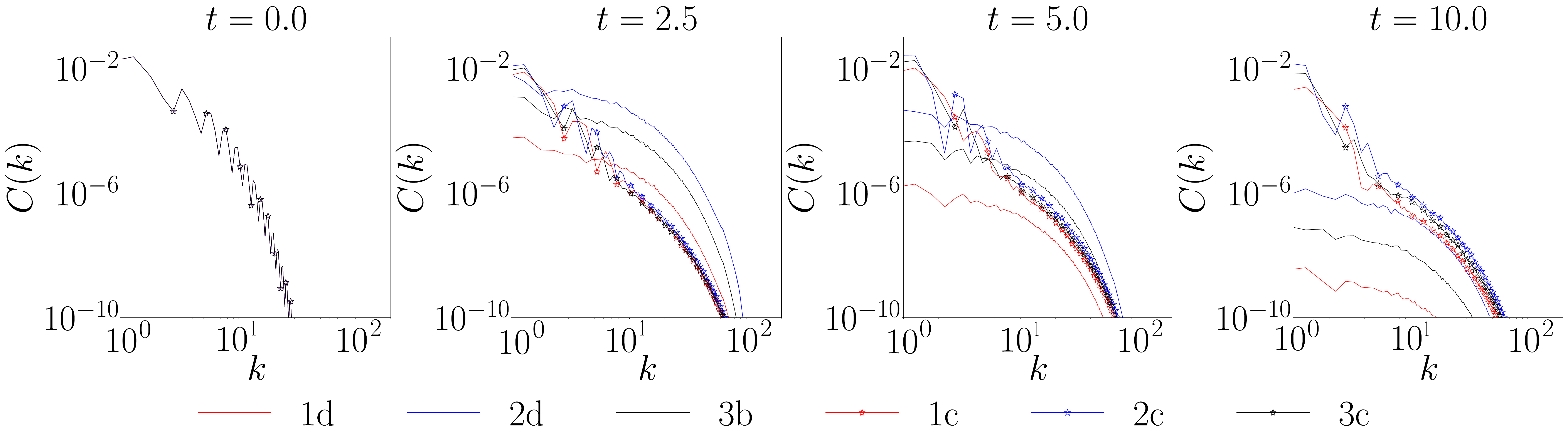}
\put(-470,125){(a)}
\put(-352,125){(b)}
\put(-235,125){(c)}
\put(-119,125){(d)}
    \caption{Scalar spectra at different time instances. The spectra broaden initially due to stirring and then decay as molecular diffusion becomes dominant. }
    \label{fig: Evolution of scalar spectrum}
\end{figure*}

\subsection{Post Stirring Behavior}

Figure \ref{fig: Grad Yc with stirring} shows that $\overline{|\nabla Y_c|}$ increases and then decreases with time. This increase is due to the increased surface area caused by corrugations. The decrease in $\overline{|\nabla Y_c|}$ occurs when stirring has exposed all the points of the spherical blob species to molecular diffusion~\cite{jossy2025mixing,meunier2003vortices,liu2022mixing}. During the stirring phase, the action of molecular diffusion can be neglected. After sufficient decrease in $\overline{|\nabla Y_c|}$, molecular diffusion becomes significant and is responsible for the homogenization of the mixture. To show the effect of diffusion, we define the scalar energy spectra as
\begin{equation}
    C(k) = |\widehat{Y_{c}}|^{2} .
\end{equation}
Figure \ref{fig: Evolution of scalar spectrum} shows the evolution of the scalar energy spectra at different time instances. At the initial times, stirring results in broadening of the spectra (figure \ref{fig: Evolution of scalar spectrum}b). This indicates that at these times, the action of molecular diffusion is minimal. In figure \ref{fig: Evolution of scalar spectrum}c, we see that the spectra decay, indicating an increase in the role of molecular diffusion. At later times, the reduction of scales in DNS cases is more than that of the synthetic cases. The spectra also show that the denser mixtures have a faster drop in scales, indicating faster homogenization.

\citeauthor{toussaint2000spectral} \cite{toussaint2000spectral} report that the spectral decay of passive scalar in chaotic mixing is exponential and self-similar. Similarly, we define reduced spectra using 
\begin{equation}
    C^{*}(k) = \overline{ C(k) / \sum_{k} C(k) },
\end{equation}
and try to fit the data as per the exponential decay given by 
\begin{align}
    &(\mathrm{Re }\mathrm{Sc})^{1/2} \tau^{*} C^{*}(k) \nonumber \\ &=  A((\mathrm{Re }\mathrm{Sc})^{-1/2} k)^{\beta} e^{-\varphi (\mathrm{Re }\mathrm{Sc})^{1/2} k }
    \label{eq: Reduced spectra},
\end{align}
where $A$, $\varphi$, and $\beta$ are dependent on the flow and the decay time $\tau^{*}$ is given by 
\begin{equation}
    \tau^{*} = \left( \mathrm{Re} \mathrm{Sc} \right) \left( \sum_k k^2 C^*_k \right)^{-1}.
    \label{eq: Time deacay}
\end{equation}

\begin{figure}[!t]
    \centering
    \includegraphics[width=1.0\linewidth]{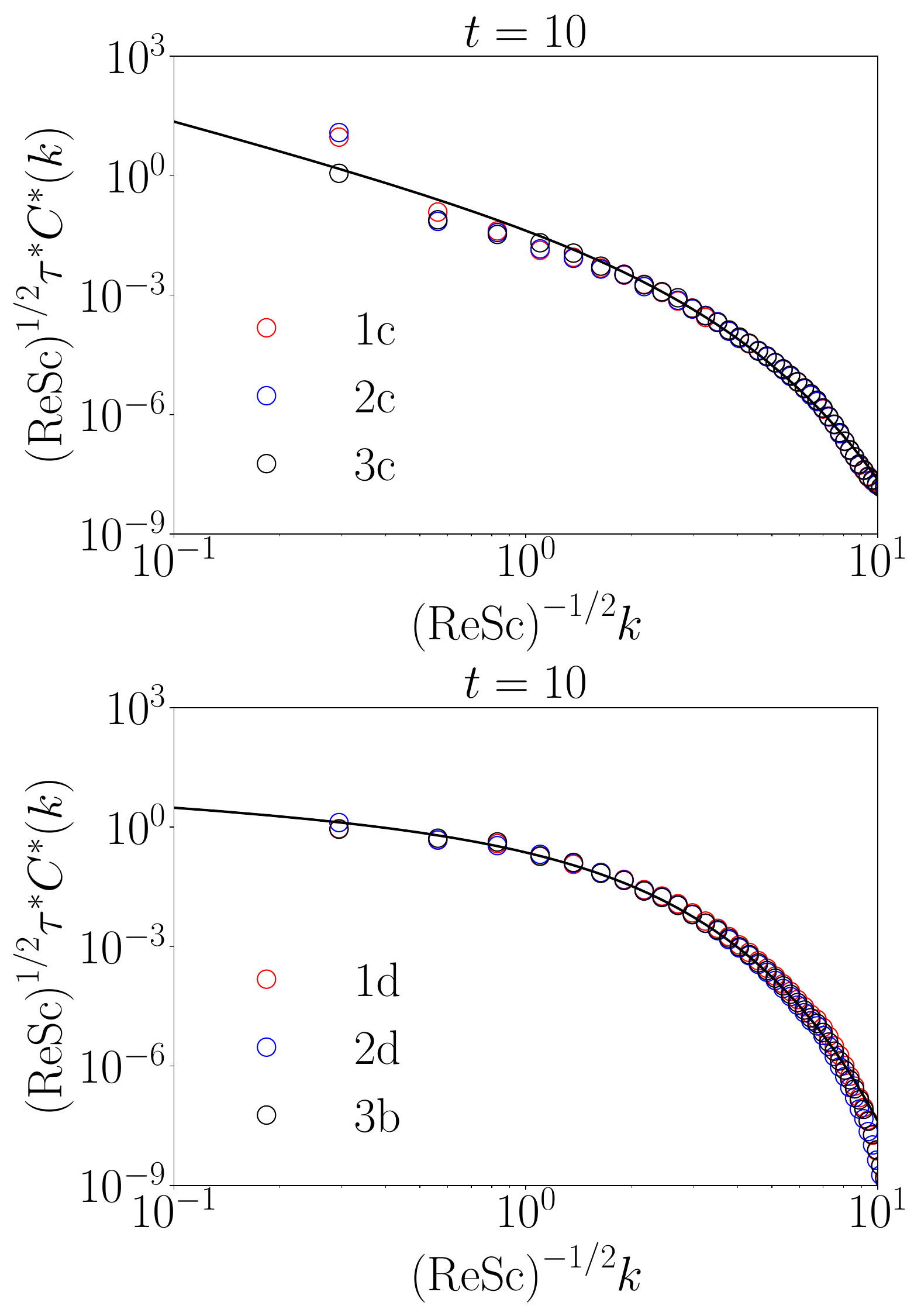}
        \put(-200,330){(a)}
    \put(-200,170){(b)}
    \caption{Reduced scalar spectra at $t=10$ for (a) synthetic cases, (b) DNS cases with same mean density $(\langle \rho \rangle) = 1$ with their corresponding exponential curves following \eqref{eq: Reduced spectra}. The active scalar energy spectrum behaves exponentially and exhibits self-similarity in the diffusion regime. For  $3\mathrm{b}$ the value of $\varphi = 1.5$ and $\beta = 0.5$, and for $3\mathrm{c}$ the value of $\varphi = 1.0$ and $\beta = 2.35$ is used to fit the data for exponential decay.}
    \label{fig: Redcued spectra}
\end{figure}

Figures \ref{fig: Redcued spectra}a, and b show the rescaled and reduced spectra of the synthetic cases and DNS cases with the same mean mixture densities, respectively. We fit the curve defined by \eqref{eq: Reduced spectra} for all the cases. In figure \ref{fig: Redcued spectra}a, we see that the synthetic cases collapse on top of each other with the fitted curve for the same curve fitting parameters. For the DNS, different parameters are required (c.f. figure \ref{fig: combined Spectra from DNS with different initial mean density}), but we see that each of the cases can be fitted as per \eqref{eq: Reduced spectra}. In  figure \ref{fig: Redcued spectra} (b) , we see that all the cases collapse on top of each other and also with the curve fit for the passive scalar case. This shows that in the diffusion-dominated regime, active scalars behave like passive scalars and undergo an exponential decay in the scalar spectra. These indicate that the effect of density gradients on the mixing process is reduced in the molecular diffusion phase.

\subsection{Comparison of DNS and synthetic turbulence}
In this section,  we compare the stirring effects of the homogeneous isotropic turbulence fields generated by Navier Stokes, and the synthetic stochastic velocity field. We see that in all cases, the trend of the denser mixture homogenizing faster remains the same. However, there is a difference in the rate of homogenization. This can be seen in figures \ref{fig: 3D figures of  Yc with stirring} and \ref{fig: 3D figures of  Yc with stirring synthetic},  and is quantified in figure \ref{fig: Evolution of Mixing quality}, where we see that the DNS cases homogenize faster. Note that initial the kinetic energy spectra for the cases labeled using ``b'' and ``c'' in table~\ref{tab: Synthetic-cases} are the same. Additionally, the kinetic energy spectra for cases labeled ``d'' and ``c'' remain same since the spatially averaged density effects are nullified. This indicates that kinetic energy spectra may not be appropriate for classifying the transient mixing dynamics.

Figures \ref{fig: 3D figures of  Yc with stirring}b and f show that the DNS cases ``toss" and ``tumble" the interface when compared to the figure \ref{fig: 3D figures of  Yc with stirring synthetic}. To quantify the effect of the spatial distribution of velocity, we calculate the area-to-volume ratio following \citeauthor{schumacher2005statistics} \cite{schumacher2005statistics}.
 This ratio indicates the degree of corrugations and folding of the interface by the stirring process. We define a threshold limit for the interface ($Y_{c0}$) and define an interface using 
 \begin{equation}
\tilde{L}_X = \{\boldsymbol{x} : \theta(\boldsymbol{x}, t) \geq Y_{c0}\}.
 \end{equation}
 We assign an indicator \( I_{ijk} = 1 \) to grid points in \(\tilde{L}_X\) and \( I_{ijk} = 0 \) otherwise, where \( i, j, k \) are grid indices ranging from 1 to \( N \). For the points falling within the threshold, the boundary points of \(\tilde{L}_X\) have 1 to 5 nearest neighbors (\( N_{ijk} \)), while inner points have 6~\cite{schumacher2005statistics}. The total number of inner points is denoted by \( \overline{N}_{ijk} \). Then the area-to-volume ratio ($\Gamma_{X}$) is given by 
\begin{equation}
    \Gamma_{X} = \frac{A_{X}}{V_{X}} = \frac{1}{\overline{N}_{ijk} \Delta x_{(i,j,k) \in \tilde{L}_X }} \sum (6 - N_{ijk}),
\end{equation}
\begin{figure}[!t]
    \centering
\includegraphics[width=1.0\linewidth]{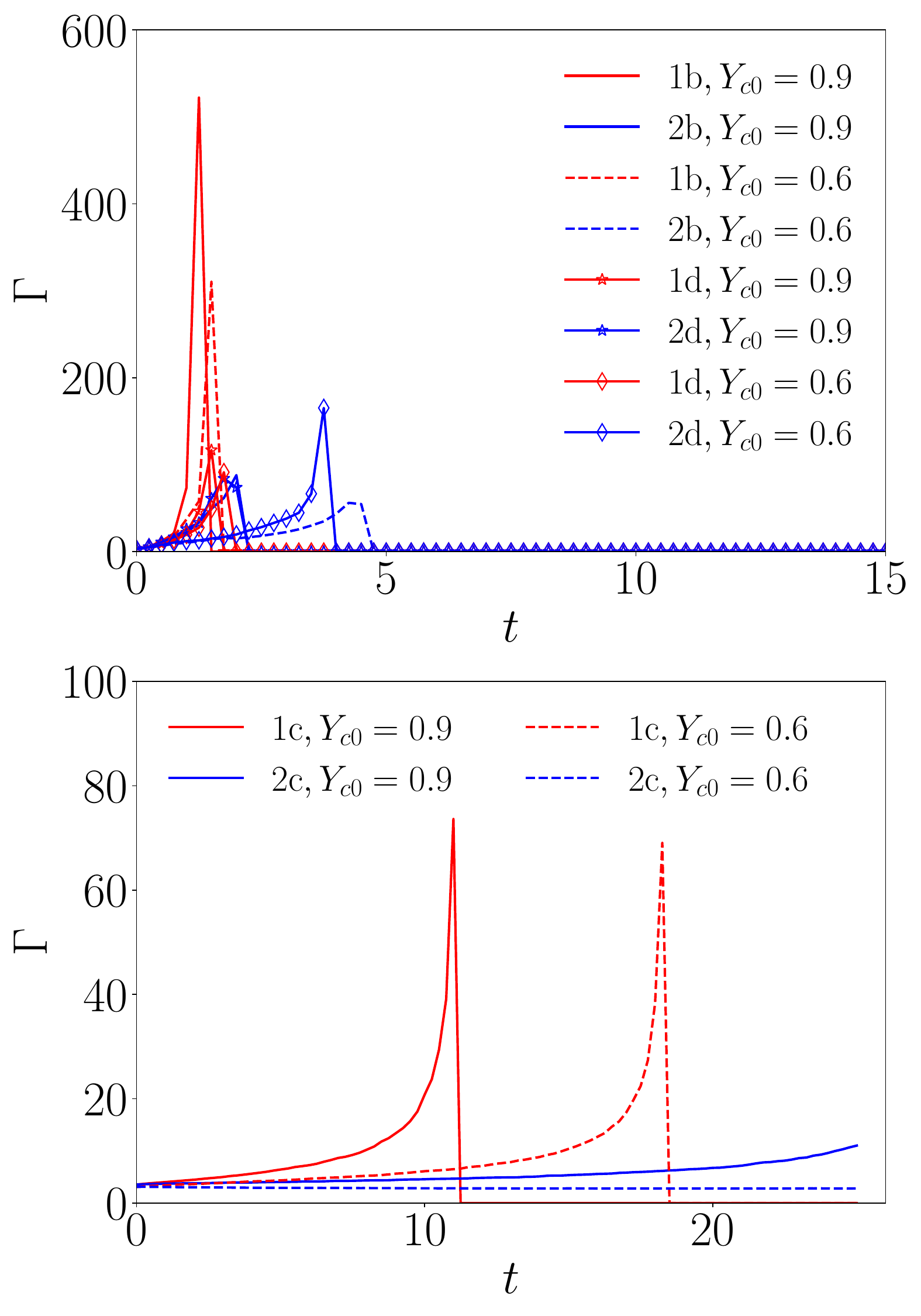}
\put(-220,330){(a)}
    \put(-220,170){(b)}
    \caption{Evolution of area-to-volume ratio for (a) DNS cases and (b) synthetic cases. The blob interface in the DNS cases undergoes more deformations than the synthetic cases.}
    \label{fig: Evolution of Area to volume ratio.}
\end{figure}
where $\Delta x$ is the grid spacing. Figures \ref{fig: Evolution of Area to volume ratio.}a and b show the evolution of the area-to-volume ratio as a function of the iso-level of concentration $Y_{c0}$ for the DNS cases (both with different $\overline{\rho}$ and same $\overline{\rho}$) and the synthetic cases, respectively. Figure \ref{fig: Evolution of Area to volume ratio.} shows that the trend of denser mixtures homogenizing faster is consistent across the type of mixing methods. This is indicated by the faster disappearance of the iso-level of concentration for the denser mixture. We also see that the $\Gamma$  is higher for the denser mixture cases when compared to the lighter mixture cases. This is true even for pure diffusion cases, where the inward movement of the interface reduces the volume faster than the surface area and thereby increases $\Gamma$ for denser mixtures. However, the value of  $\Gamma$  is higher for the DNS cases when compared to the synthetic turbulence cases. This indicates that more surface area is exposed to the action of molecular diffusion, allowing for faster homogenization. The increase in surface area is one of the characteristics of stirring. This shows that for the same spectra, the spherical blob in the DNS cases undergoes more stretching and folding. This affects the rate of homogenization. In the synthetic turbulence cases, the turbulence structures are created through the spatially filtered white noise. However, no interactions exist among them in time, nor is there any convection of the formed turbulent structures. \Revision{}{The synthetic turbulence cases do not account for the nonlinear term of the Navier-Stokes equation \cite{marti1997langevin} nor the intermittent effects ( the Gaussian properties properties are preserved due to the linear nature of the synthetic field). The current study indicates that these could be  essential for studying the mixing evolution characteristics} . The lack of these features results in a lesser ``tossing" and ``tumbling" of the spherical blob by the synthetic turbulence cases. These indicate that the choice of mixing method used for the study is important for analyzing the mixing dynamics.

 \section{Conclusions} 
\label{sec: Conclusions}
We have studied the mixing of active scalars by stochastic velocity fields using three-dimensional turbulence generated from the Navier Stokes equation and synthetically generated stochastic field. We consider the mixing of two different species of varying densities where one of the species is concentrated in a spherical blob. The Atwood number is used as the dimensionless parameter to study the effect of the stirring on the mixing of a denser mixture, lighter blob $(At > 0)$ and a lighter mixture, denser blob $(At < 0)$. The presence of density gradients alters the mixing characteristics of the scalars. In the absence of stirring, we show that a heavier blob expands outward in a lighter surrounding and a lighter blob shrinks inward in a heavier surrounding. We define a mixture as a denser mixture or a heavier mixture if the surrounding species is denser than the blob, and vice-versa.  A denser mixture is found to homogenize faster than a lighter mixture. In pure diffusion cases, the denser and lighter mixture can be simplified as an unsteady diffusion equation of density with two different initial conditions. In the presence of stirring, we see that the rate of homogenization trends is faster with different mixing methods than in the pure diffusion case. We also show that the stirring enhances the diffusion process where the diffusion coefficient is augmented by the velocity correlations. The Navier-Stokes generated turbulence field is found to homogenize the mixtures faster than the synthetically generated stochastic fields. We use the mixing variance $\sigma^2$ to analyze the rate of homogenization. We show that, unlike passive scalars, the spreading of the active scalars does not necessarily indicate better mixing. However, once the diffusion starts dominating the stirring process, active and passive scalars behave identically. We show that in the diffusion regime, the scalar energy spectrum is self-similar and decays exponentially. We also show that for low mach turbulence, the effect of species density gradients on the velocity field is minimal in the kinetic energy spectra.

On comparison of the mixing methods, we find that the Navier Stokes generated turbulence field folds and stretches the interface of the two species more than the synthetic stochastic field case. We show that kinetic energy spectra, which have been used as a standard in a lot of previous studies \cite{lacasta1995phase,sokolov2000bimolecular,guseva2017aggregation,sharma2019two,pant2021mixing} for characterizing synthetic turbulence-driven mixing, is not the only quantity that should match the actual turbulent field. \Revision{}{The synthetic cases are unable to account for the intermittent effects of turbulence, which could have an effect on the mixing dynamics. We defer the explicit role of intermittency on mixing to a future study. }  Additionally, the convection of larger eddies (time correlation of larger length scales) are not captured in the synthetic fields, resulting in negligible stretching and folding of the inhomogeneity and a slower mixing dynamics. The current study highlights that better modeling of synthetic turbulence is necessary for studying both passive scalar and active scalar mixing. The current study also indicates that denser and lighter mixtures require different stirring processes for optimized and faster mixing. 

\subsubsection*{Acknowledgements.}
We acknowledge the financial support received from Science and Engineering Research Board (SERB), Government of India under Grant No. SRG/2022/000728 and the financial support received from the Ministry of Ports, Shipping, and Waterways, Government of India under Grant No. ST-14011/52/2021-MT(349089). We also thank IIT Delhi HPC facility for computational resources.


  \bibliographystyle{elsarticle-num-names} 
  \bibliography{references}



\end{document}